\documentclass[apj]{emulateapj}
\usepackage{graphicx}

\newcommand\one{{\sc i}}
\newcommand\ii{{\sc ii}}
\newcommand\iii{{\sc iii}}

\shorttitle{Resolving the FIR Line Deficit}
\shortauthors{Croxall et al.}

\begin{document}  

\title{Resolving the Far-IR Line Deficit: Photoelectric Heating and Far-IR Line Cooling in NGC~1097 and NGC~4559\thanks{Herschel is an ESA space observatory with science instruments provided by European-led Principal Investigator consortia and with important participation from NASA.}}
\author{Kevin V. Croxall$^1$, J.D. Smith$^1$, M.~G. Wolfire$^2$, H. Roussel$^3$, 
K. M. Sandstrom$^4,19$, B. T. Draine$^5$, G. Aniano$^5$, D.~A. Dale$^6$, L. Armus$^7$, P. Beir\~{a}o$^7$, G. Helou$^8$,  A. D. Bolatto$^2$,
P.~N. Appleton$^8$, B.~R. Brandl$^9$, D. Calzetti$^{10}$, A.~F. Crocker$^{10}$, M. Galametz$^{11}$, B. A. Groves$^{4}$, C.-N. Hao$^{12}$, L. K. Hunt$^{13}$, B. D. Johnson$^{14}$, R. C. Kennicutt$^{11}$, J. Koda$^{15}$, O. Krause$^6$, 
Y. Li$^{10}$, S. E. Meidt$^4$, E. J. Murphy$^16$, N. Rahman$^2$, H.-W. Rix$^4$, M. Sauvage$^{17}$, E. Schinnerer$^4$, F. Walter$^4$, C. D. Wilson$^{18}$}
\affil{$^1$Department of Physics \& Astronomy, University of Toledo, 2801 W Bancroft St, Toledo, OH 43606
\\$^2$Department of Astronomy, University of Maryland, College Park, MD 20742, USA
\\$^3$Institut dÕAstrophysique de Paris, UMR7095 CNRS, Universit\'{e} Pierre \& Marie Curie, 98 bis Boulevard Arago, 75014 Paris, France
\\$^4$Max-Planck-Institut f¬ur Astronomie, K\"{o}nigstuhl 17, D-69117 Heidelberg, Germany
\\$^5$Department of Astrophysical Sciences, Princeton University, Princeton, NJ 08544, USA
\\$^6$Department of Physics \& Astronomy, University of Wyoming, Laramie, WY 82071, USA
\\$^7$Spitzer Science Center, California Institute of Technology, MC 314-6, Pasadena, CA 91125, USA
\\$^8$NASA Herschel Science Center, IPAC, California Institute of Technology, Pasadena, CA 91125, USA
\\$^9$Leiden Observatory, Leiden University, P.O. Box 9513, 2300 RA Leiden, The Netherlands
\\$^{10}$Department of Astronomy, University of Massachusetts, Amherst, MA 01003, USA
\\$^{11}$Institute of Astronomy, University of Cambridge, Madingley Road, Cambridge CB3 0HA, UK
\\$^{12}$Tianjin Astrophysics Center, Tianjin Normal University, Tianjin 300387, China
\\$^{13}$INAF - Osservatorio Astrofisico di Arcetri, Largo E. Fermi 5, 50125 Firenze, Italy
\\$^{14}$Institut dÕAstrophysique de Paris, UMR7095 CNRS, Universit\'{e} Pierre \& Marie Curie, 98 bis Boulevard Arago, 75014 Paris, France
\\$^{15}$Department of Physics and Astronomy, SUNY Stony Brook, Stony Brook, NY 11794-3800, USA
\\$^{16}$Observatories of the Carnegie Institution for Science, 813 Santa Barbara Street, Pasadena, CA 91101, USA
\\$^{17}$CEA/DSM/DAPNIA/Service dÕAstrophysique, UMR AIM, CE Saclay, 91191 Gif sur Yvette Cedex, France
\\$^{18}$Department of Physics \& Astronomy, McMaster University, Hamilton, Ontario L8S 4M1, Canada
\\$^{19}$Marie Curie Fellow
}
\email{kevin.croxall@utoledo.edu}

\begin{abstract}
The physical state of interstellar gas and dust is dependent on the processes which heat and cool this medium.  To probe heating and cooling of the ISM over a large range of infrared surface brightness, on sub-kiloparsec scales, we employ line maps of [C~\ii] 158 $\mu$m, [O~\one] 63 $\mu$m, and [N~\ii] 122 $\mu$m in NGC~1097 and NGC~4559, obtained with the PACS spectrometer onboard {\it Herschel}.  We matched new observations to existing Spitzer-IRS data that trace the total emission of polycyclic aromatic hydrocarbons (PAHs).  We confirm at small scales in these galaxies that the canonical measure of photoelectric heating efficiency, ([C~\ii] + [O~\one])/TIR, decreases as the far-infrared color, $\nu f_\nu$(70 $\mu$m)/$\nu f_\nu$(100 $\mu$m), increases.  In contrast, the ratio of far-infrared (far-IR) cooling to total PAH emission, ([C~\ii] + [O~\one])/PAH, is a near constant $\sim$6\% over a wide range of far-infrared color, 0.5 \textless\ $\nu f_\nu$(70 $\mu$m)/$\nu f_\nu$(100 $\mu$m) $\lesssim$ 0.95.  In the warmest regions, where $\nu f_\nu$(70 $\mu$m)/$\nu f_\nu$(100 $\mu$m) $\gtrsim$ 0.95, the ratio ([C~\ii] + [O~\one])/PAH drops rapidly to 4\%.  We derived representative values of the local UV radiation density, $G_0$, and the gas density, $n_H$, by comparing our observations to models of photodissociation regions. The ratio $G_0/n_H$, derived from fine-structure lines, is found to correlate with the mean dust-weighted starlight intensity, $\langle U\rangle$ derived from models of the IR SED.  Emission from regions that exhibit a line deficit is characterized by an intense radiation field, indicating that small grains are susceptible to ionization effects.  We note that there is a shift in the 7.7 / 11.3 $\mu$m PAH ratio in regions that exhibit a deficit in ([C~\ii] + [O~\one])/PAH, suggesting that small grains are ionized in these environments.
\end{abstract} 
 
 \keywords{galaxies: individual (NGC 1097, NGC 4559) --- galaxies: ISM --- ISM: lines and bands}
 
 \section{Introduction}
The life of gas and dust in galaxies is largely cyclic in nature. Interstellar gas and dust collapse to form stars, stars in turn process the material via nuclear fusion and, at the end of their lives, return a portion of this processed gas back into the interstellar medium (ISM), thereby altering and enriching the gas to be used in the formation of the next generation of stars.    One key to mediating this process is the heating and cooling cycle of interstellar gas reservoirs.  The physical conditions of interstellar clouds determine if those clouds collapse and form additional stars.  If gas is unable to cool sufficiently, star formation is quenched.  Thus, the efficiency of heating and cooling cycles in neutral gas will affect the evolution of the gas and dust in the ISM.

Neutral gas is thought to be heated primarily through photoelectrons that are ejected from dust grains by incident UV radiation \citep[e.g.,][]{watson,TH85,wolfire1995,wolfire2003}.  One of the key contributors of photoelectrons is thought to be the nearly ubiquitous polycyclic aromatic hydrocarbons (PAHs) that emit strongly in the mid-infrared (IR) \citep[e.g.,][]{bakes,weing}.  Small particles such as PAHs are excited by UV photons and are subsequently radiatively de-excited \citep{draine07}.  The vibrational bands emitted by PAHs should thus indirectly trace the ejection of photoelectrons, which are thought to dominate the heating of the neutral ISM.  Ejected photoelectrons may heat a wide range of environments from dense molecular gas (T $\sim$ 10 - 50 K, $n_H$ $\sim$ 10$^3$ - 10$^6$ cm$^{-3}$) to diffuse warm neutral gas (T $\sim$ 5000 K, $n_H$ $\sim$ 0.6 cm$^{-3}$) \citep{drainebook}.

On the other side of the balance between heating and cooling are the far-IR fine-structure lines that cool the heated gas.  As it is relatively easy to excite ($\Delta$E/k $\sim$ 92 K), all but the coldest gas is cooled by the collisionally excited [C~\ii] 158 $\mu$m line.  The abundance of carbon and the low ionization potential (11.26 eV) results in the [C~\ii] fine-structure line at 158 $\mu$m being one of the brightest emission lines seen in normal star forming galaxies \citep{luhman}.  This line emission escapes the gas and dust allowing the gas to cool and thus provides a measure of the gas heating rate, assuming that the ISM has reached a local equilibrium.  In addition to [C~\ii] emission, the IR portion of the spectrum also hosts other key fine-structure lines (e.g.,  [N \ii] 122 $\mu$m, [O~\one] 63 $\mu$m, [Si \ii] 35 $\mu$m, etc.).  This confluence of fine-structure lines and the PAH complexes in the mid- to far-IR makes this spectral range especially effective at mapping the heating and cooling of the ISM.

\citet{malhotra} examined globally averaged emission from far-IR fine-structure lines in a sample of 60 star forming galaxies as part of the ISO Key Project Survey of Normal Galaxies.  They showed that there is a decrease of line emission relative to continuum emission in the far-IR in galaxies with warmer far-IR colors.  This trend has persisted as investigations have measured emission from C$^+$ in more than 180 unresolved galaxies \citep{brauher}.  Mid-IR studies have also shown that the luminosity of PAHs decreases relative to the luminosity of larger dust grains, rather than remain constant with dust temperature, when looking at warmer far-IR colors \citep{helou1991}. This decline of PAH emission relative to large grains has been explained by invoking intense radiation fields that ionize PAHs \citep{tielens1999}.  Alternatively, the depressed ratio of PAH emission to total infrared surface brightness (TIR) could be due to sampling different mixtures of gas phases within the large ($\sim$70\arcsec) ISO beam \citep{helou}.  Recently, the unprecedented sensitivity of  {\it Herschel} has enabled the detection line deficits in all far-IR line fine structure lines \citep{Carpio}.  This has been interpreted as an increased competition for UV photons in dusty galaxies, as traced by the ratio of the far-IR continuum and the mass of molecular gas observed in a galaxy.  Thus, significant progress has been made in understanding the heating and cooling of the interstellar gas of nearby galaxies.  However, these efforts have been hampered by limited spatial resolution, spectral coverage, and sensitivity.  

With the Photodetector Array Camera \& Spectrometer (PACS) onboard the {\it Herschel Space Observatory} we are for the first time able to target the dominant cooling lines of neutral and ionized interstellar gas over sub-kiloparsec scales in nearby galaxies.  The {\sc Kingfish} project \citep[Key Insights on Nearby Galaxies: a Far- Infrared Survey with Herschel, ][]{kingfish} is an open-time Herschel key program investigating critical heating and cooling diagnostics of dust and gas within nearby galaxies. In particular, {\sc Kingfish} targets fine-structure lines that are the principal coolants  of  predominantly neutral gas ([C~\ii] 158 $\mu$m and [O~\one] 63 $\mu$m) and strong diagnostic lines of ionized gas ([O \iii] 88 $\mu$m and [N~\ii] 122 $\mu$m) in nearby galaxies that span a wide breadth of key physical parameters \citep{kingfish}.  To complete the picture of heating and cooling of the ISM in nearby galaxies, we combine the far-IR observations of the {\sc Kingfish} program with mid-IR observations available from the Spitzer Infrared Nearby Galaxies Survey \citep[{\sc Sings},][]{sings}.  The spectral coverage and spatial resolution of these combined observations allow us to investigate the distribution of PAH features and far-IR line emission on scales previously restricted to the Milky Way and Magellanic Clouds.

\begin{figure*}[htp] 
\epsscale{1.2}
   \centering
   \plotone{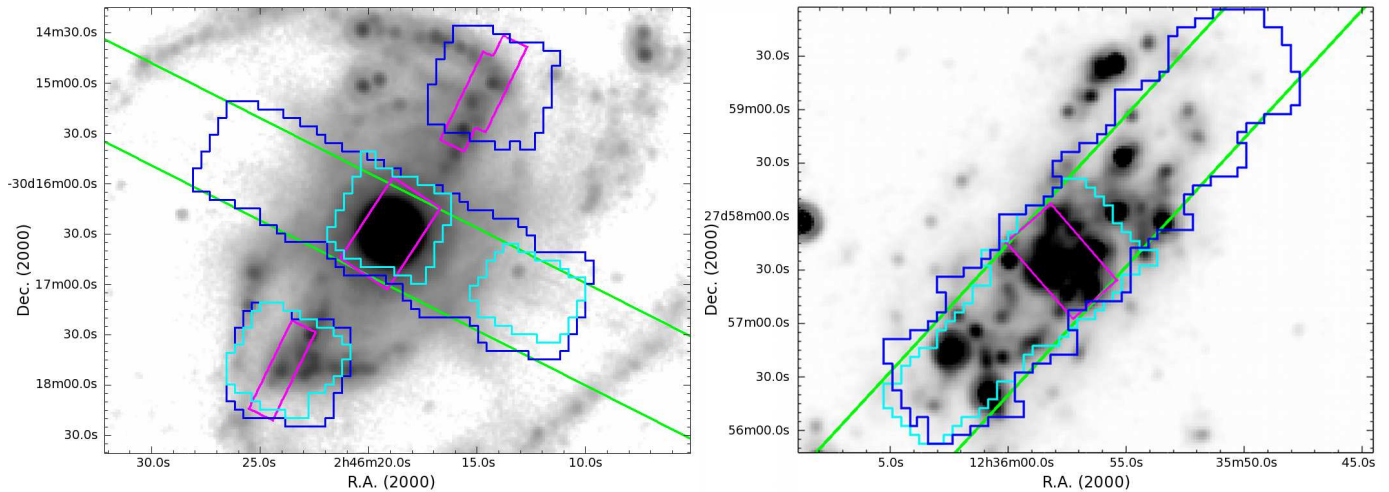}
   \caption{Observational footprints for IR spectral observations of NGC~1097 (left) and NGC~4559 (right) overlaid on MIPS 24$\mu$m images.  The green box denotes the area which had full Long-Low coverage with IRS.  The magenta regions mark the area observed in Short-Low with Spitzer-IRS.  The PACS-WS observational footprint is shown in Blue.  PACS-CN coverage is shown in cyan.}
   \label{fig:aors}
\end{figure*}   

NGC~1097 and NGC~4559 were selected as {\sc Kingfish} Science Demonstration Phase (SDP) targets.  NGC~4559 is a late spiral type galaxy (SABcd) at a distance of approximately 8.45 Mpc.  It has super-solar metallicity\footnote{Compared to the solar oxygen abundance of \citet{asplund}, 12+log(O/H) = 8.69$\pm$0.05.} (12+log(O/H) = 8.81\footnote{We have adopted for this work the abundances derived using the theoretical calibration of \citet{KK04}.  Adoption of the empirical calibration of \citet{PT05} would yield subsolar oxygen abundances for NGC~1097 and NGC~4559.  This choice slightly affects the magnitude, but not the direction, of the difference in metallicity between NGC~4559 and NGC~1097.}) and shows no signs of AGN activity \citep{Moustakas}.  In contrast, NGC~1097 is slightly more enriched than NGC~4559 (12+log(O/H) = 9.09) and is a Seyfert 1 galaxy that contains a bright central ring of star formation approximately 2 kpc in diameter \citep{Moustakas,Hummel,sandstrom}.  At a distance of 19.1 Mpc \citep{willick}, the central PACS pointing of NGC~1097 is one of the brightest regions in the {\sc Kingfish} sample, when ranked by average TIR surface brightness.  

These two systems  clearly probe a wide range of environments: intense, concentrated star formation in the ring in NGC~1097 to nominal, distributed star formation in the disks of both galaxies.   This range of environments gives a broad spread in far-IR color space allowing a robust baseline over which we may study the heating efficiency.  Indeed, the {\sc Kingfish} program has been designed to sample the widest possible range of star forming environments found in the nearby Universe.  While NGC~1097 and NGC~4559 represent only a small fraction of the entire {\sc Kingfish} sample, they span a representative range of surface brightness and radiation fields for an exploratory study of the heating and cooling mechanisms operating in the ISMs of nearby galaxies. 

In this paper we present a broad investigation and comparison of properties of the ISM in NGC~1097 and NGC~4559 by combining data from the {\sc Kingfish} and {\sc Sings} programs.  For a detailed investigation of heating and cooling in the central ring and ends of the bar in NGC~1097 see also \citet{beirao2011}.  In \S2 we present the observations and discuss the data processing.  As both chopped and un-chopped PACS spectral data were acquired, these data are compared to each other as well.  To accurately trace total PAH emission despite varying spectral coverage, we derive a method to combine limited spectral and photometric data sets in \S3.  In \S4 we discuss the heating and cooling of gas in these galaxies.  The far-IR line deficit is investigated across the full range of star-forming environments in these two galaxies using the two strongest coolants, the [C~\ii] 158 $\mu$m and [O~\one] 63 $\mu$m lines.  We discuss how the cooling is related to observed PAH emission in the ISM and investigate scenarios that have been proposed to explain the far-IR line deficit in \S5.  Finally, we summarize the results of this study in \S6. 

\section {Observations}
\subsection{Acquisition and Processing of PACS Data}
Using PACS onboard Herschel we carried out far-IR spectral observations of NGC~1097 and NGC~4559 as part of the {\sc Kingfish} SDP program.  We report a detailed summary of observations in Tables \ref{t:obs1097} and \ref{t:obs4559}.  The integral field unit (IFU) of PACS has a roughly 47\arcsec$\times$47\arcsec\ field of view composed of 25 9\farcs4 $\times$ 9\farcs4 pixels \citep{pacs}.  In Figure \ref{fig:aors} we show footprints of the spectral observations of NGC~1097 and NGC~4559.  These are aligned with Spitzer Infrared Spectrograph \citep[IRS,][]{houck} Long-Low (LL) coverage.  The long strips measure  240\arcsec $\times$ 60\arcsec\ and  290\arcsec $\times$ 60\arcsec\ for NGC~1097 and NGC~4559 respectively.  In NGC~4559, observations are aligned with the major axis; whereas in NGC~1097 the LL strip is nearly orthogonal to the star forming bar.  The position angle of observations was determined by constraints in the pointing of the Spitzer telescope.  In this work we highlight resolved mapping observations obtained for three key emission lines ([C~\ii] 158$\mu$m, [N~\ii] 122$\mu$m,  and [O I] 63$\mu$m) in both Chop-Nod (CN) and Wavelength Switching\footnote{This now deprecated mode was superseded by the unchopped mode of PACS spectroscopy.} (WS) modes (see Tables \ref{t:obs1097} and \ref{t:obs4559}).

All PACS spectral observations were reduced using the Herschel Interactive Processing Environment (HIPE) version 8.0 \citep{hipe}.  Reductions applied the standard spectral response functions, flat field corrections, and flagged instrument artifacts and bad pixels  \citep[see][]{pacs}.  A dark was applied only to observations taken in WS mode, as CN observations remove the dark current via chopping.  As inflight flux calibrations are not yet available for PACS spectral data, ground based calibrations were applied to the data.  These calibrations result in absolute uncertainties on the order of 30\% and relative uncertainties of 10\% and 20\% in the blue and red grating orders respectively.  

Wavelength switching data were offered as an alternative to chopping when a clean chop position was not available. In WS mode the grating is dithered by approximately half the FWHM of an unresolved line profile in an AABBBBAA pattern, where A is the reference grating position and B is the grating position of interest.  The flux at the grating position of interest may be compared to the flux detected at the reference position to remove the baseline, rather than using a reference at the same grating position with a spatial displacement.  Observations of an off position before and after each observation supplement the spectral dithering.  Following the standard reduction procedures for PACS, as outlined by \citet{pacs}, we would subtract signal detected in the reference grating position from the grating position of interest.  As the grating dither is small compared to the line width, this subtraction creates a differential line profile that must be modeled to reconstruct the data obtained in WS mode.  

Fortunately, because of the exceptional thermal stability of PACS, we were able to process WS mode observations without modeling the differential line profile.  Instead, observations were processed under the assumption that the baseline was stable for the duration of the observation.  This meant that WS reductions were not dependent upon models of line profiles to measure the total flux.  However, due to drifts and long term transients, the astrophysical continuum was undetected.  In essence, this reduction process was the same as that used for the Unchopped mode that has since replaced WS in post SDP observations.  

The averages of the clean off observations obtained were subtracted from observations to correct for the thermal background contributed by {\it Herschel}.  Subsequently, all spectra within a given spatial element were combined after offsetting each to the median continuum value.  Approximately one fifth of the measured ramps exhibited a highly unphysical continuum resulting from transients in the detector.  These transients were strongly correlated with motions of the grating and of {\it Herschel} itself.  Pixels which deviated from the mean spectrum by more than five standard deviations were masked out.  Finally, spectral cubes with 5\arcsec\ spatial pixels were created by projecting dithered raster positions of each spectral line together.  In CN data, the two nod positions of the PACS array are slightly rotated on the sky with respect to one another.  Therefore the nods are treated as separate raster positions when projecting the observations.

To optimize the signal to noise and still take advantage of the spatial resolution, we employed 10\arcsec\ apertures tiled every 5\arcsec\ along the observed footprint.  This aperture corresponds to $\sim$930~pc at the distance of NGC~1097 and $\sim$410~pc at the distance of NGC~4559.  As apertures are not completely independent of their neighbors, there will be correlation between adjacent measurements.  However, this approach avoids introducing bias into the sample by selecting apertures by eye.  Furthermore, it fills out as large a parameter space as possible in the environment sampled for this study.  

\begin{figure}[bp] 
   \centering
   \epsscale{1.2}
   \plotone{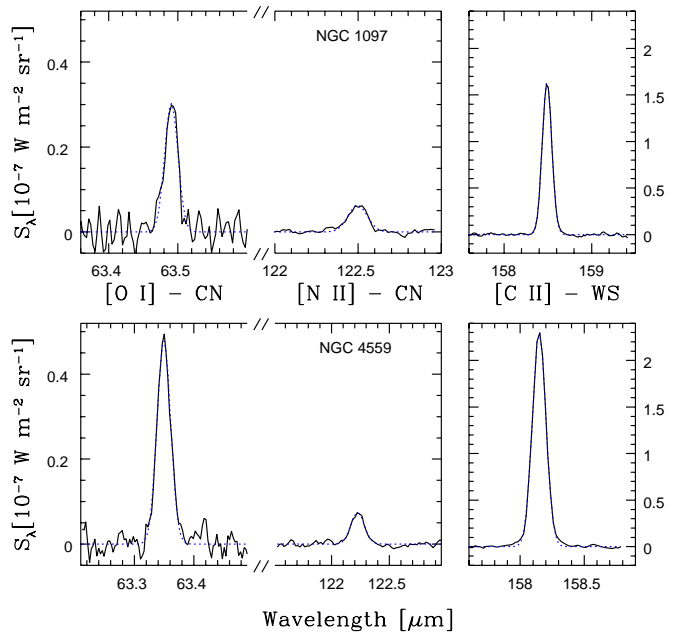}
   \caption{Representative continuum subtracted spectra from regions with [N~\ii] 122$\mu$m detections in diffuse regions of NGC~1097 (top) and the nuclear region of NGC~4559 (bottom).  The [O~\one] 63$\mu$m line and the [N~\ii] 122$\mu$m line are plotted on the same scale.  All spectra shown were extracted from the chop-nod observations.  Dotted blue lines show Gaussian fits to each line.  For spectra from the ring of NGC~1097 we refer the reader to \citet{beirao2010}.}
   \label{fig:spectra}
\end{figure}

Some apertures exhibited a residual baseline after projection. To remove this, we fit a first order polynomial to the spectral range outside one FWHM on either side of the line center as estimated from HI observations of NGC~1097 and NGC~4559 \citep{Koribalski,tully}; this baseline fit to the spectrum was then subtracted.  Subsequently, line fluxes for each aperture were determined from PACS observations using two methods.  We measured the flux by directly integrating under the continuum subtracted line. Additionally, we used Gaussian profiles fit to the spectrum extracted from each aperture to determine the flux (see Figure \ref{fig:spectra}).  For sources with detections above the three sigma level, we found that the flux measurements from both methods agreed within the relative uncertainties (20\%) in NGC~4559.  Lines from the central region of NGC~1097 are not well described by a Gaussian profile due to the large velocity dispersion of the star forming ring (see Figure \ref{fig:spectra}).  Thus, the two flux determinations differed significantly in this spatial region.  We have therefore adopted fluxes determined from direct integration.  Outside the ring, both methods of integration over a line are in agreement. 

\begin{figure*}[htp] 
   \centering
   \epsscale{1.2}
   \plotone{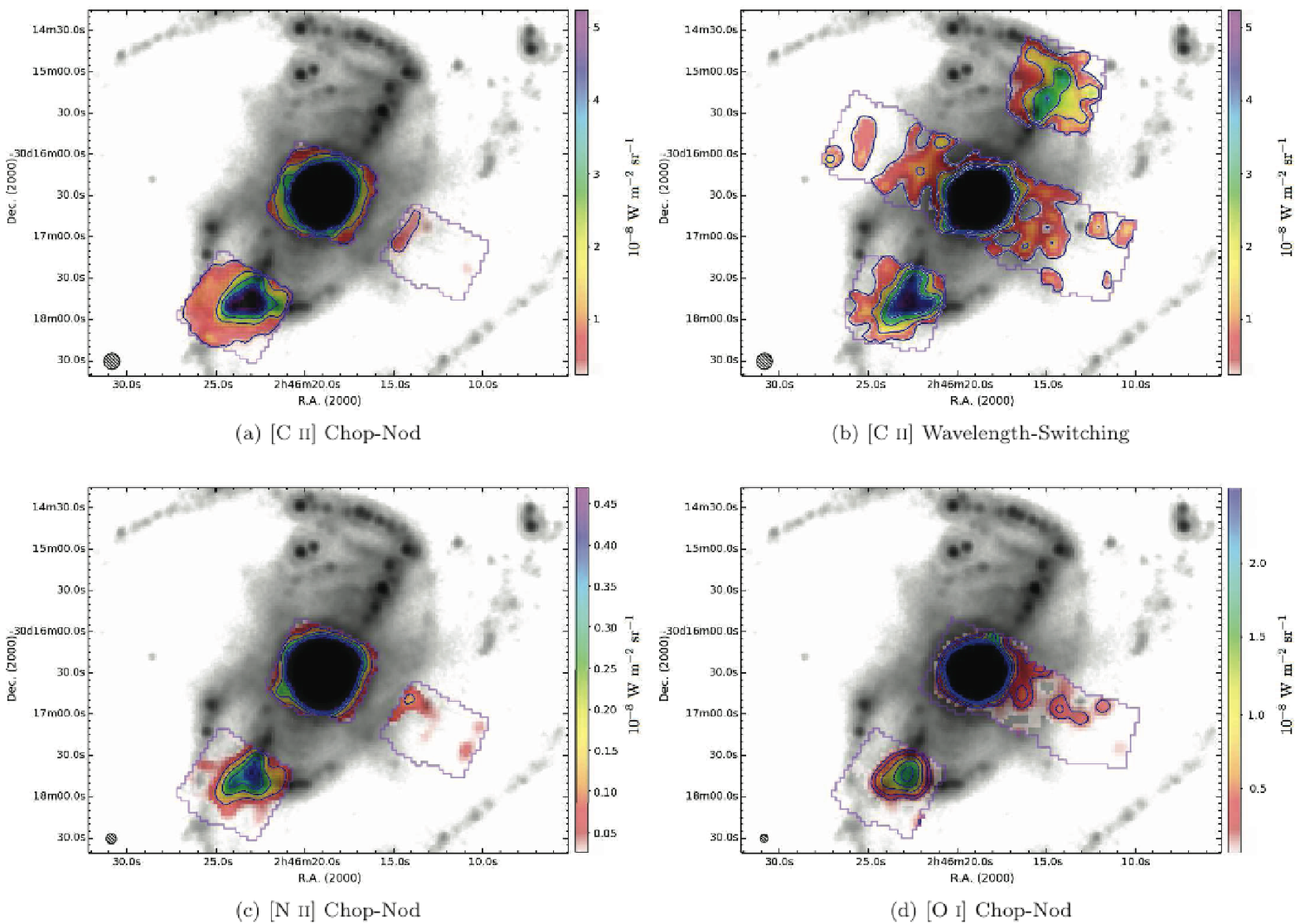}
     \caption{Contour maps of far-IR fine-structure lines in NGC~1097 overlaid on the MIPS 24 $\mu$m image.  The footprint of the spectral mapping in each line is shown as a faint purple outline.  PACS beam sizes at the appropriate wavelengths are shown at the lower left of each panel.  Contours begin at a SNR of 3$\sigma$; [C~\ii] contours are plotted every 1.2 $\times$ $10^{-9}$W m$^{-2}$ sr$^{-1}$, until 4 $\times$ $10^{-8}$W m$^{-2}$ sr$^{-1}$; [N~\ii] contours are plotted every 1 $\times$ $10^{-9}$W m$^{-2}$ sr$^{-1}$ until 4; [O~\one] contours are plotted every 2 $\times$ $10^{-9}$W m$^{-2}$ sr$^{-1}$ until 1.4 $\times$ $10^{-8}$W m$^{-2}$ sr$^{-1}$.  Line maps have been scaled to show the diffuse emission.  For line maps of nuclear emission we refer the reader to \citet{beirao2010}}
   \label{fig:linemapsa}
\end{figure*}

\begin{figure*}[htp] 
   \centering
   \epsscale{1.2}
   \plotone{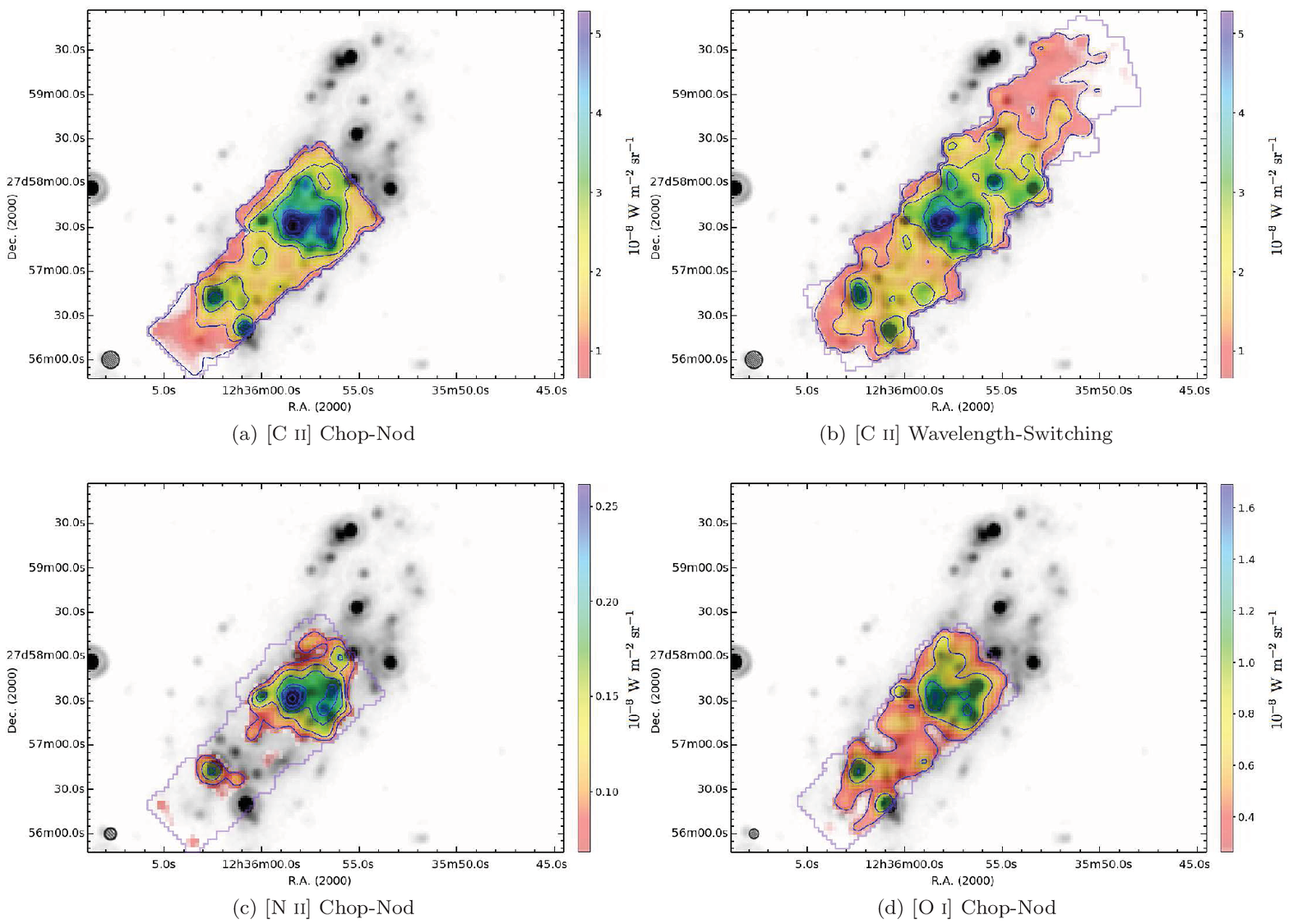}
     \caption{Contour maps of far-IR fine-structure lines in NGC~4559 overlaid on the MIPS 24 $\mu$m image.  The footprint of the spectral mapping in each line is shown as a faint purple outline.  PACS beam sizes at the appropriate wavelengths are shown at the lower left of each panel.  Contours begin at a SNR of 3$\sigma$; [C~\ii] contours are plotted every 1 $\times$ $10^{-8}$W m$^{-2}$ sr$^{-1}$; [N~\ii] contours are plotted every 4 $\times$ $10^{-10}$W m$^{-2}$ sr$^{-1}$; [O~\one] contours are plotted every 4 $\times$ $10^{-9}$W m$^{-2}$ sr$^{-1}$.}
   \label{fig:linemapsb}
\end{figure*}

\begin{figure}[htb] 
   \centering
   \epsscale{1.2}
   \plotone{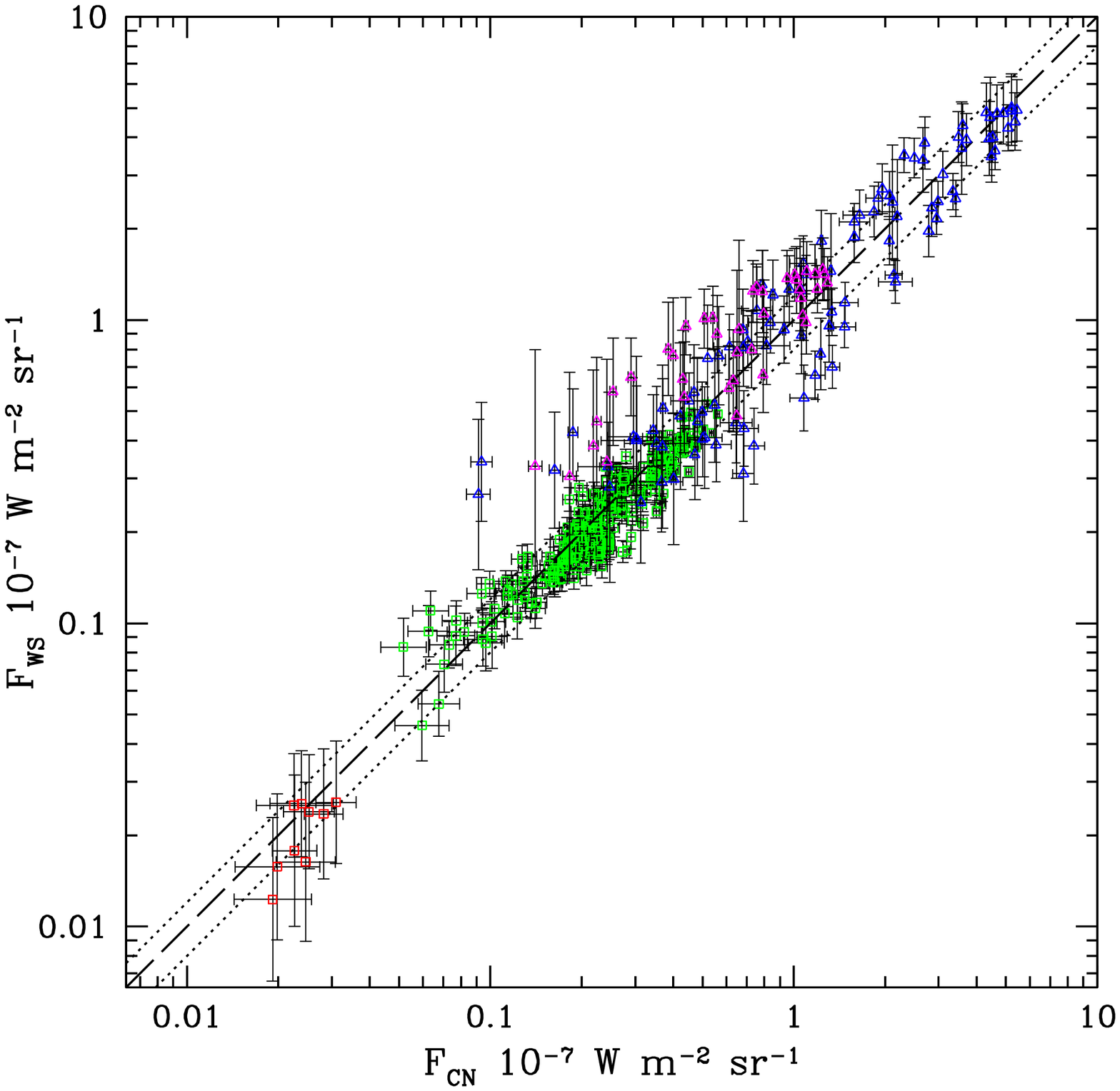}
   \caption{Flux measurements from data acquired in WS mode are plotted against the flux measurements from CN data.  Line fluxes were measured from spectra extracted from 10\arcsec\ square WCS matched apertures.  Measurements of the [C~\ii] 158$\mu$m line are plotted as blue triangles and green squares, for NGC~1097 and NGC~4559 respectively; measurements of the [N~\ii] 122$\mu$m line are plotted as magenta triangles and red squares, for NGC~1097 and NGC~4559 respectively.  The solid line is a one-to-one correlation; dotted lines are indicative of the 20\% relative uncertainty of PACS due to calibration uncertainty.}
   \label{fig:cn2ws}
\end{figure}

The PACS spectral observations of NGC~1097 and NGC~4559 are shown as projected line maps in Figures \ref{fig:linemapsa} and \ref{fig:linemapsb}.  Line maps of NGC~1097 have 3$\sigma$ detection thresholds at a surface brightness of $\sim$ 0.04, 0.07, 0.01, and 0.07 $\times$ $10^{-7}$ W m$^{-2}$ sr$^{-1}$ in [C~\ii] CN, [C~\ii] WS, [N~\ii] CN, and [O~\one] CN lines respectively.  Line maps of NGC~4559 have 3$\sigma$ detection thresholds at a surface brightness of $\sim$ 0.03, 0.05, 0.01, and 0.03 $\times$ $10^{-7}$ W m$^{-2}$ sr$^{-1}$ in [C~\ii] CN, [C~\ii] WS, [N~\ii] CN, and [O~\one] CN lines respectively.  While the [N~\ii] 122 $\mu$m and [O~\one] 63 $\mu$m lines were only detected in roughly 50\% of PACS pixels, even short integrations of the [C~\ii] 158 $\mu$m line yielded detections in 80\% of the PACS pixels.  Indeed, only when observations were taken off the axis of NGC~1097 was [C~\ii] detection limited.\\

\subsection{Supplementary Data}
Both galaxies were also observed by the {\sc Sings} program \citep{sings} with Spitzer-IRS \citep{houck}.  The central regions of NGC~1097 and NGC~4559  were targeted by the short-low (SL) module. The two extra-nuclear regions targeted in NGC~1097 were also observed with SL.  In addition to SL observations, the long-low (LL) module targeted strips across each target galaxy (see Figure \ref{fig:aors}).  There was no overlap of LL observations with the extra-nuclear positions in NGC~1097.   For a more detailed description of the Spitzer data, including reduction techniques and a presentation of the central spectra of NGC~1097 and NGC~4559 we refer the reader to \citet{pahfit}.  For examples of the small aperture IRS spectra and maps of PAH emission we refer the reader to \citet{beirao2011}

Spitzer-IRS spectra were extracted using CUBISM \citep{cubism}.  Broad spectral features and the mid-IR lines were fit and decomposed using PAHFIT \citep{pahfit}.   We fixed the silicate absorption optical depth at zero as \citet{pahfit} found that negligible amounts of absorption from silicate dust were needed to reproduce the equivalent widths of lines in kpc scale  regions of IRS spectral observations of NGC~1097 and NGC~4559.  Only data that had a surface brightness larger than three times the root mean square (RMS) noise were accepted as physically meaningful detections.    

The PACS and IRS spectral observations are supplemented by imaging from 3 - 160 $\mu$m obtained as part of the {\sc Kingfish} and {\sc Sings} programs.  These images are essential to the determination of the total infrared continuum emitted by dust, the far-IR color, and to aid in accounting for PAH emission where full spectral coverage is not available.  The reduction of observations followed standard procedures, as detailed by \citet{sandstrom} and \citet{dalesingssed}.  

\subsection{Convolution of Data}
We convolve data to ensure that the comparison of fluxes data obtained at different resolutions is valid. This is particularly important as we are using a variety of instruments to cover a large wavelength range (3 -- 160 $\mu$m).  We are also interested in resolving the signatures of heating and cooling on small scales.  These interests must be balanced as they lead us in diametrically opposite directions in the treatment of the data.

Convolving to a larger point-spread function (PSF) will degrade the resolution of our data, reducing the dynamic range of heating properties being sampled.  This is further complicated by the fact that data near the edge of a map cannot be properly convolved as there is a hard edge past which no data is obtained.  Therefore, at the edges of a map the flux is not conserved.  While map edges lie well beyond the regions of interest in our imaging data, this is not the case for the spectral maps.  

To test where we are limited in convolving the data, we have convolved masks of our IRS spectral maps (e.g., the value of each pixel recording data is set to one and pixels outside the footprint are set to zero) to the largest beam size used in this study (PACS PSF at 160 $\mu$m, FWHM = 11\farcs01) using sTinyTim \citep{krist}.  We found this effect to be modest when convolving Spitzer-IRS LL spectra.  Regions 22\arcsec\ wide at the center of LL strips (see Figure \ref{fig:aors}) conserve approximately 98\% of the flux, before dropping to 70\% of the appropriate flux at the edge.  However, the Spitzer-IRS SL spectra data suffer a substantial loss of flux; only 90\% of the flux is conserved at the center of the spectral maps, dropping to 30-50\% at the edges.  Furthermore, \citet[][see Appendix A]{Pereira-Santaella} show that the PSF of Spitzer-IRS spectral maps are not well reproduced by modeled telescope PSFs, suggesting that the generation of maps has significantly altered the PSF.

As a compromise between matching resolution and retaining reasonable spatial coverage without losing flux, we have convolved data with a beam smaller than the PACS 100 $\mu$m band (FWHM = 6\farcs97) to this resolution and bootstrapped data with a beam larger than the PACS 100 $\mu$m band down to this resolution.  Convolution kernels from \citet{aniano} were used to convolve IRAC 3.6 $\mu$m (FWHM = 1\farcs67), IRAC 8.0 $\mu$m (FWHM = 2\farcs03), MIPS 24 $\mu$m (FWHM = 6\farcs52), and PACS 70 $\mu$m (FWHM = 5\farcs41) images to the resolution of the PACS 100 $\mu$m band (FWHM = 6\farcs97).  As the delivered PSFs of Spitzer-IRS are extremely complicated due to the mapping strategy employed by SINGS \citep{Pereira-Santaella}, we have adopted a single Gaussian kernel appropriate for a central wavelength in each IRS module.  Convolution of the [O~\one] 63$\mu$m data (FWHM $\sim$ 5.3\arcsec) was performed assuming a Gaussian PSF.  

To overcome the resolution limits for PACS 160 $\mu$m (FWHM = 11\farcs01), we convolve PACS 70 $\mu$m observations to the resolution of PACS 160 $\mu$m.  A representative 70 $\mu$m / 160 $\mu$m ratio was measured in 12\arcsec\ apertures centered on the associated spectral aperture.  Similar to the approach in \citet{pahfit}, we assume that this color ratio is the same in both the 12\arcsec\ and 10\arcsec\ apertures.  This ratio is combined with the photometry from the PACS 70 $\mu$m observations that are convolved to the resolution of the PACS 100 $\mu$m band to derive a surface brightness for PACS 160 $\mu$m at this resolution.  We calculate the TIR following the prescription of \citet{draine}, 
\begin{equation}TIR = 0.95\langle\nu S_\nu\rangle_8 + 1.15\langle\nu S_\nu\rangle_{24} + \langle\nu S_\nu\rangle_{70} + \langle\nu S_\nu\rangle_{160}\end{equation}
using the matched 24 $\mu$m and 70 $\mu$m photometry  combined with the bootstrapped PACS 160 $\mu$m, where the S$_{\nu}$ are the surface brightness values in the indicated bandpasses.  While the [C~\ii] 158 $\mu$m line falls in the PACS 160 $\mu$m bandpass, contamination from this line is less than 2\%, therefore we do not correct for its inclusion in the calculation of TIR.

Following our approach with the PACS 160 $\mu$m band described above, we have bootstrapped the [C~\ii] data to the resolution of the PACS 100$\mu$m band.  Here we assume that the ratio of [C~\ii]/$\nu f_\nu(70)$ does not change between 12\arcsec\ apertures, which sample the full 160 $\mu$m beam, and the 10\arcsec\ apertures.  This ratio was  combined with PACS 70 $\mu$m observations that were convolved to the resolution of the PACS 100 $\mu$m band to derive a [C~\ii] surface brightness in the 10\arcsec\ apertures.  Bootstrapping  [C~\ii]/TIR and [C~\ii]/PAH in this manner alters their surface brightness by $\lesssim$5\%.  The most significant changes occur in the warmest regions of NGC~4559, which lie at the south-west edge of the [C~\ii] map next to a bright point source.  The increased coverage of the 12\arcsec\ aperture is not completely contained by the [C~\ii] map, thereby skewing the measured ratio of [C~\ii]/$\nu f_\nu(70)$.

To ensure that the conclusions of this study are not affected by convolving to data to the PACS 100 $\mu$m band rather than the 160 $\mu$m band, we have also convolved all our data to the resolution of the PACS 160 $\mu$m band. Using the coarser resolution and 12\arcsec\ apertures to ensure the PACS 160$\mu$m beam is fully sampled, we reproduce the trends presented in this paper; however, the coarser resolution and larger apertures greatly reduce the number of reliable apertures.

\subsection{Comparison of CN and WS Data}
Standard spectral observations with the PACS spectrometer require chopping between the target and a blank space of sky at a high frequency to remove the thermal signatures of the telescope and to correct for transients.  The PACS chopper is limited to a maximum throw of 6\arcmin.  When targeting many nearby galaxies, an offset of 6\arcmin\ is insufficient to avoid emission from the target galaxy in the offset field.  NGC~1097 and NGC~4559 have minor axis diameters of  6\arcmin.3  and 4\arcmin.4 respectively.  These sizes provide clean off positions within the 6\arcmin\ throw of the telescope chopper, given the orientation of the  observations.  Therefore, these galaxies provided an opportunity to verify that un-chopped data can reproduce the line flux measurements from standard chop-nod observations.  

 Line data acquired in WS mode have $\sim$40\% less wavelength coverage than CN data due to the decreased number of grating steps taken in the WS observations.  The increased spectral resolution of the blue channel results in insufficient wavelength coverage to ascertain an accurate placement of the continuum.  Therefore, we restrict our comparison of these two modes, CN and WS, to data acquired from the red photoconductor array, which has lower spectral resolution.  Spectral windows were centered on the nominal velocity of each galaxy.  At the center of NGC~1097 the line-of-sight velocity differs by $\sim$500~km s$^{-1}$ on the northern and southern edges of the rotating inclined star forming ring.  This leads to wide double peaked lines in many of the PACS spectral pixels.  This rotational broadening of the line results in an increased uncertainty in the bright [C~\ii] line as even less continuum is available for accurate placement of the baseline.  
 
 [C~\ii] 158$\mu$m and [N~\ii] 122$\mu$m fluxes measured from CN observations are plotted against matched observations in Figure \ref{fig:cn2ws}.  CN and WS data clearly agree within the $\sim$20\% relative uncertainty of the flux calibrations.  Observations of the [N~\ii] 122 $\mu$m line in NGC~1097 show the largest deviations from unity.  This is most certainly due to the combination of the intrinsic faintness of the [N~\ii] line and the uncertainty of the continuum placement in WS mode in the broad lines at the center of NGC~1097.

While both CN and WS modes prove to be sensitive to detections of emission lines, we adopt CN data when available due to the (1) increased wavelength coverage that enables a robust placement of the continuum and (2) the increased depth of the CN observations of [N~\ii] and [O~\one] which, in many cases, had integration times that were larger than the comparable WS observation by a factor of two or more.  As both observation modes provide suitable line detections, we make use of the [C~\ii] WS data to extend the spatial coverage of both galaxies.    

\section{Tracing PAH emission}
Fully sampling the total PAH emission requires coverage by both the LL and SL modules of the IRS.  However, a significant portion of our PACS coverage area is observed by only a single IRS module: LL 15-40 \micron\ in the strips, SL 5-15 \micron\ in the extra-nuclear regions.  To expand our measurement of small grain heating efficiency to all regions with detections of [C~\ii] and [O~\one] it is necessary to deduce the total PAH power via available data.  

To estimate the total PAH power with limited spectral coverage, we use the limited areas in which full 5-38 \micron\ coverage of PAH emission exists in NGC~1097 and NGC~4559 to formulate measures of the strength of aromatic bands for apertures that have either LL or SL coverage, but not both.  We define the total PAH power as the combined intensity of dust features tabulated by \citet[][see Table 3]{pahfit}.  These features were measured in the complete combined IRS spectra.  We selected major PAH emission complexes in LL and SL that are representative of the overall PAH emission to trace the total PAH power in regions where observations of only one module were carried out.  We chose the 17 \micron\ complex in the LL spectra to represent total PAH emission as this is the strongest feature covered by the LL module (14-38 \micron) \citep[e.g., ][]{smith04,werner}.  We select the 11.3 \micron\ complex as representative of PAH emission in the SL module (5-14 \micron).  The 11.3 \micron\ complex is neither the only nor the strongest PAH feature observed by SL.  We selected this feature as it was a clean band that was subject to less uncertainty in the placement of the continuum, given the relatively weak silicate absorption features in these objects.  This is a requisite attribute of a representative PAH feature in SL as the uncertainty in the placement of the continuum underlying these bands is increased if LL data are not fit along with SL spectra.  

In addition to spectral measures of PAH, the strongest PAH band complex is included in the IRAC 8.0 \micron\ bandpass.  To increase the precision of using photometric data to determine PAH emission, IRAC 3.6 \micron\ data are used to subtract the stellar contribution to the 8.0\micron\ band \citep[as in][]{helou04}.  We correct the 8.0 \micron\ IRAC band for the stellar component by scaling, and then subtracting, the 3.6 \micron\ IRAC images, as determined from interpolating stellar SEDs for galaxies in the Local Volume Legacy survey \citep{marble}.  
\begin{equation}PAH^*_{Phot} ~=~ [\nu S_\nu(8.0) - 0.24\times \nu S_\nu(3.6)].\end{equation} 
We note that this prescription leads PAH$^*_{Phot}$ to be $\sim$10\% larger than the total power in PAH features.  This may be due in part to the fact that we have not removed the warm dust continuum which is also present in the 8.0 \micron\ IRAC images.  However, it may also indicate that the stellar correction given by \citet{marble} is on the low side \citep{helou04,Stefano11}.

\begin{figure}[tbp] 
   \centering
   \epsscale{1.2}
   \plotone{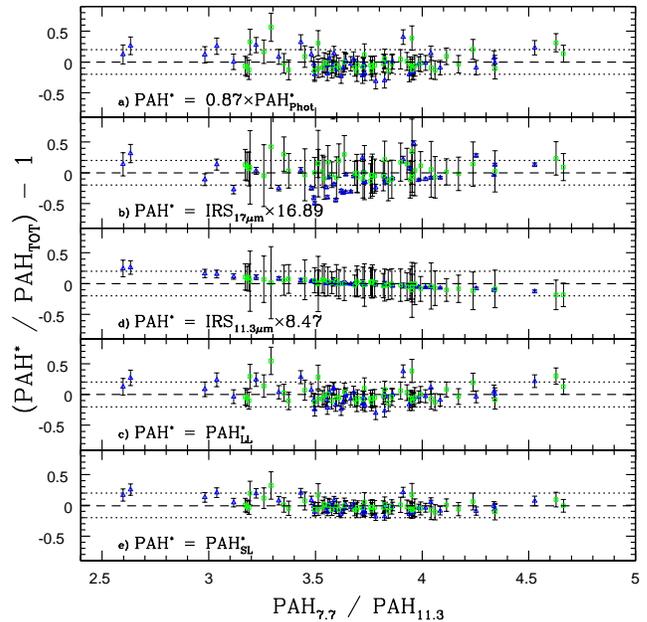}
   \caption{We plot the ratio of five estimators of the total PAH power and the true PAH power measured from the full (5-38 \micron) IRS spectrum as a function of the 7/11.3 \micron\ PAH ratio in regions where full 5--38\micron\ spectra are available, e.g., the nuclear regions of NGC~1097 and NGC~4559.}
\label{fig:pahstarl}
\end{figure}

The coefficients of linear combinations of photometrically and spectroscopically measured PAH bands were adjusted to minimize the RMS scatter.  
As seen in Figure \ref{fig:pahstarl} where we plot the ratio of the estimated PAH power (PAH$^*$) and the true spectroscopic PAH power ($PAH_{TOT}$), the estimated total PAH$^*$ is found to be
 \begin{equation}PAH^*_{LL} ~=~ 0.7472\times[PAH^*_{Phot} + 3.25\times \nu S_\nu(17)],\end{equation} 
 and 
 \begin{equation}PAH^*_{SL} ~=~ 0.497\times[PAH^*_{Phot} + 3.59\times \nu S_\nu(11.3)],\end{equation} 
using either LL or SL data as well as IRAC data, where available.  We note that PAH$^*_{SL}$ and PAH$^*_{LL}$ are unbiased tracers of the total PAH emission and are robust to changes in the small grain population, e.g., the ionization state of the PAHs or changes to the grain size distribution, unlike individual PAH bands.    For instance, a clear trend with PAH ionization exists when the 11.3 \micron\ PAH feature alone is used to ascertain the total PAH power, see the middle panel in Figure \ref{fig:pahstarl}.
     
Regions near the edges of our IRS maps suffered a loss of flux in the convolution of Spitzer-IRS cubes. To retain these apertures in our sample, we measured both $PAH^*_{Phot}$ and the appropriate spectroscopic PAH, either $PAH^*_{LL}$, $PAH^*_{SL}$, or $PAH_{TOT}$.  A fraction of $PAH^*_{Phot}$ equivalent to the fraction of lost flux, as measured from convolved masks of IRS modules, was added to the appropriate spectroscopic PAH measurement. This scaling resulted in a $\lesssim$5\% correction in the total PAH power attributed to these regions.

We find the 1~$\sigma$ scatter with respect to the mean is 0.16, 0.21, 0.10,  0.15 and 0.09 dex for PAH$^*_{Phot}$,IRS$_{17}$, IRS$_{11.3}$, PAH$^*_{LL}$ and PAH$^*_{SL}$, respectively.  Thus, using both the spectral and photometric information together to measure the total power in PAH emission significantly reduces the scatter when only LL is available.  While there is not a significant change in the scatter between IRS$_{11.3}$ and PAH$^*_{SL}$, the combined data show much less dependence on the ionization state of the PAHs.  We can further reduce the scatter in PAH$^*_{SL}$ by including all the main PAH features present, however, this increases the correlation between PAH$^*_{SL}$ and the ratio of the 7.7/11/3 \micron\ bands and is strongly dependent on the placement of the continuum, which is uncertain without LL observations.

\section{Heating the ISM}
The efficiency of heating interstellar gas is controlled by the size and ionization state of the small dust grains that emit photoelectrons.  The photoelectric efficiency is defined as the ratio of gas heating (via photoelectrons) to total dust heating (via ultraviolet and optical photons)\footnote{This definition of the heating efficiency is based on observable properties and therefore differs from the definition given by \citet{TH85}, i.e., the ratio of energy that goes into gas heating compared to the far-ultraviolet radiation absorbed by small grains and PAHs.}, 
\begin{equation} \epsilon_{\mbox{{\tiny PE}}} = \frac{\Gamma_{\rm{gas}}}{\Gamma_{\rm{dust}}},\end{equation}
\citep{mochi}.  Dust heating and gas cooling can be observed through emission in the infrared.  Thermal radiation from radiatively-warmed dust peaks in the far-IR while the neutral gas, heated by electrons liberated from small dust grains, cools via collisionally excited fine-structure lines.  The most prominent of the cooling lines are the [C~\ii] 158 \micron, [O~\one] 63 \micron\ lines which dominate in low \citep{wolfire1995,wolfire2003} and high density \citep{kaufman99,kaufman06} environments respectively.  These two lines together constitute the bulk of the cooling in neutral gas.  Indeed, the third most luminous neutral gas line, [Si \ii] 35 \micron, contributes roughly an order of magnitude less to the cooling rate than both [C~\ii] 158 \micron\ and [O~\one] 63 \micron\ individually \citep{kaufman06}.  

Observationally, it is assumed that the far-IR line-to-continuum ratio traces $\epsilon_{\mbox{{\tiny PE}}}$ such that 
\begin{equation}\frac{I_{\mbox{[C~\ii]}} + I_{\mbox{[O~\one]}}}{TIR} = (\eta_{\mbox{[C~\ii]}} + \eta_{\mbox{[O~\one]}})\epsilon_{\mbox{{\tiny PE}}},\end{equation} 
where $\eta$ is the fractional contribution of each line to the total gas cooling, and ($\eta_{\mbox{[C~\ii]}}$ + $\eta_{\mbox{[O~\one]}}$) $\simeq$ 1.  This implicitly assumes contributions from other cooling sources (e.g., [C~\one], [Si~\ii], CO, etc.) are negligible.  Under this premise, \citet{malhotra} demonstrated that the global value of $\epsilon_{\mbox{{\tiny PE}}}$ decreases in galaxies that exhibit a warmer far-IR color, F$_\nu(60~\mu$m)/F$_\nu(100~\mu$m).  This drop, or deficit, in $\epsilon_{\mbox{{\tiny PE}}}$ can be attributed to a decrease in the rate of photoelectric heating, $\Gamma_{\rm{gas}}$, as dust temperature and  star formation intensity increase.  As the local UV radiation density, $G_0$, measured in units of the local average interstellar field, increases relative to the particle density, $n_H$, the effects of grain charging gain in importance.  Therefore a common explanation for the decrease of $\epsilon_{\mbox{{\tiny PE}}}$ has been the ionization of small grains \citep[e.g.,][]{malhotra}.  Photoelectric heating becomes less efficient as grains are ionized, resulting in a decrease in cooling line emission relative to the warm dust continuum.  

However, some studies have indicated that a high $G_0$ alone cannot produce the range of line ratios observed in galaxies that exhibit a line deficit.   \citet{negishi} found $G_0/n_H$ to be independent of far-IR color, indicating the mechanism producing the line deficit is related to the gas density.  Alternatively, an excess amount of dust in the ionized gas could add to the far-IR continuum while suppressing [C~\ii] emission, leading to the deficit \citep{luhman2,abel2009}.  Indeed, recent results from {\it Herschel} showed that all far-IR fine-structure lines, independent of whether the line originates in ionized or neutral gas, exhibited a deficit relative to FIR at high values of L$_{FIR}$/M$_{H_2}$ \citep{Carpio}; these results were apparent in a variety of galaxy types (e.g., LINERs, Seyferts, H~\ii\ galaxies, and high-z galaxies).  Yet it is not clear that the line deficit is independent of galaxy type.  Still, other studies indicate that variations in PAH abundance can have significant effects on the the efficiency of heating and cooling in the ISM \citep{habart}.  Intense radiation fields are found in regions where PAHs are destroyed \citep{rubin}, thus the existence of a line deficit in regions with a warm far-IR color may be tracing changes in the PAH abundances.


Although great strides have been made in understanding heating and cooling in nearby galaxies with the best techniques and spatial resolution available \citep{brauher,rubin}, the use of global values can make it difficult to ensure that emission averaged together arises from similar physical environments.  \citet{contursi} noted that the integrated far-IR emission measured in global studies appears to be dictated by relatively few active regions that dominate the signal within the beam, whereas \citet{mizutani} showed that overlapping photodissociation regions (PDRs) within an aperture lead to discrepant far-IR diagnostics.  It has also been shown that the observed [C~\ii] line deficit is increased if emission that does not originate from the dense gas around massive stars (e.g., non-PDR emission) is present \citep{luhman2}.  Given our sub-kiloparsec spatial resolution, we can confidently delineate regions dominated by diffuse and dense phases, although there will inevitably be some mixture of phases even at this resolution.

\begin{figure*}[htp] 
   \centering
   \plotone{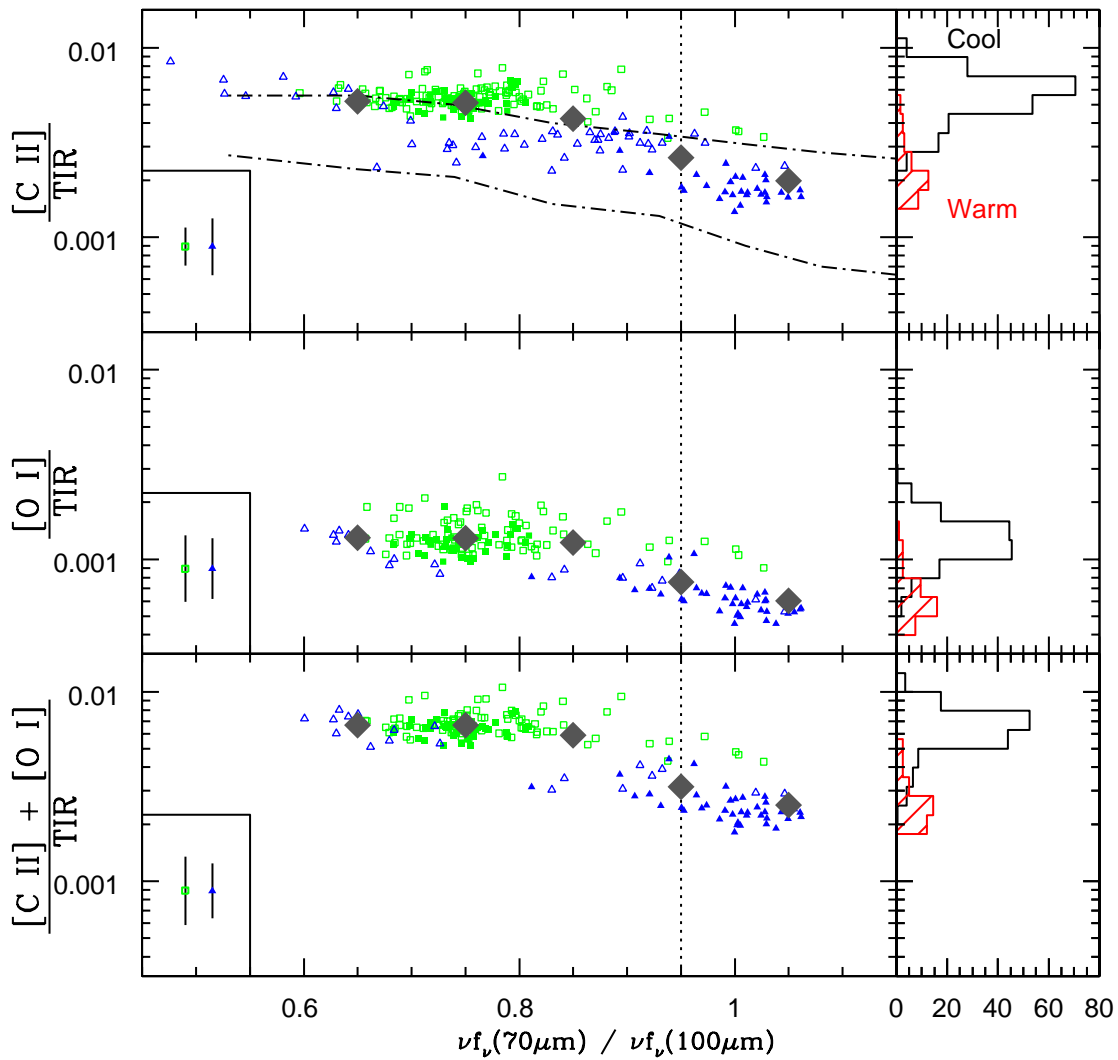}
   \caption{Top: [C~\ii] 158 $\mu$m / TIR plotted as a function of the far infrared color, $\nu f_\nu(70) / \nu f_\nu(100)$.  Middle: [O~\one] 63 $\mu$m / TIR plotted as a function of the far infrared color, $\nu f_\nu(70) / \nu f_\nu(100)$.  Bottom: ([C~\ii] 158 $\mu$m + [O~\one] 63 $\mu$m) / TIR plotted as a function of the far infrared color, $\nu f_\nu(70) / \nu f_\nu(100)$.  Even when accounting for the oxygen emission from denser regions, the photoelectric heating efficiency decreases with increasingly intense radiation. Regions in NGC~1097 are plotted as blue triangles  whereas regions from NGC~4559 are plotted as green squares.  The large grey diamonds show the weighted mean when data are binned. Histograms show the distribution of heating intensity for regions with gas warmer (red) and cooler (black) than $\nu f_\nu(70) / \nu f_\nu(100)$=0.95.  The dash-dot lines in the top panel represent 80\% inclusion curves of the galaxies from the ISO study of \citet{malhotra}.  We remind the user that there exists a 30\% uncertainty in the absolute calibration of PACS data when comparing these curves to the data. The mean error for each data set are shown to the lower left. Filled points indicate regions wherein the full IRS spectrum has been obtained.}
   \label{fig:mal}
\end{figure*}

In Figure \ref{fig:mal} we plot the ratio of strong cooling lines to TIR\footnote{Most investigations of the far-IR line deficit have used FIR band (42-122$\mu$m) as a proxy for the IR continuum rather than TIR (3-1100$\mu$m).  Use of this band does not change these trends, as TIR/FIR $\sim$ 2.} as a function of PACS 70/100 $\mu$m emission; histograms on the right of Figure \ref{fig:mal} show the distribution of $\epsilon_{\mbox{{\tiny PE}}}$ among regions cooler (black) and warmer (red) than $\nu f_\nu(70) / \nu f_\nu(100)$=0.95 (see \S5).  We use the ratio $\nu f_\nu(70) / \nu f_\nu(100)$ to trace the dust temperature as it is observationally easy to determine and enables a more direct comparison to previous studies of the far-IR line deficit.  We note that $\nu f_\nu(70) / \nu f_\nu(100)$ cannot be readily converted into $G_0$ due to contributions of stochastically heated grains to the 70 $\mu$m flux \citep{ingalls2001}.  While a robust measure of the radiation field requires detailed fitting of the spectral energy distribution \citep[e.g.,][]{draine}, we find $\nu f_\nu(70) / \nu f_\nu(100)$ is well correlated with the starlight intensity heating obtained from SED models, see \citet{aniano2} and \citet{dalekf}.

 Over a significant portion of both NGC~1097 and NGC~4559, we find that emission from the two most important cooling lines, [C~\ii] 158 $\mu$m and [O~\one] 63 $\mu$m, respectively account for an average of 0.50$\pm$0.01\%, 0.13$\pm$0.05\% of TIR in cool dust regions and 0.22$\pm$0.01\% and 0.07$\pm$0.04\% of TIR in warm dust environments.  Similar to previous studies of global measurements \citep[e.g., ][]{luhman,malhotra,Carpio}, we find that the $\epsilon_{\mbox{{\tiny PE}}}$ of resolved regions of the ISM in NGC~1097 and NGC~4559 drops in areas characterized by warmer IR colors.  This agreement is highlighted by the dash-dot lines shown in the top panel of Figure \ref{fig:mal} that delineate the parameter space filled by nearby galaxies in \citet{malhotra}.  

Even with the inclusion of [O~\one], the dominant coolant for dense environments, we find that the $\epsilon_{\mbox{{\tiny PE}}}$ is decreased in environments with more intense heating (see the bottom panel in Figure \ref{fig:mal}).  It is obvious that $\epsilon_{\mbox{{\tiny PE}}}$ is not uniform throughout a galaxy.  Rather, local conditions substantially alter the processes that heat and cool the ISM.  This is in agreement with \citet{rubin} who found that, compared to diffuse gas, regions of intense star formation in the LMC (e.g., 30 Doradus) exhibit a suppressed $\epsilon_{\mbox{{\tiny PE}}}$.

\citet{brauher} showed that $\epsilon_{\mbox{{\tiny PE}}}$ tends to decrease as a function of the far-IR color.  This decline of line emission relative to TIR is readily evident in regions within NGC~1097 whereas NGC~4559 is remarkably uniform.  While NGC~4559 does not exhibit a strong decreasing trend with far-IR color, this is likely due to the paucity of regions with warm dust that are observed.  We also note that an offset appears to exist between these galaxies that increases the measured dispersion in the line-to-TIR ratio.  As the $\epsilon_{\mbox{{\tiny PE}}}$ of a gas cloud depends on the physical conditions of the gas and its location, we can conclude that the trend seen in \citet{malhotra} and \citet{brauher} is not simply the result of mixing different phases of the ISM on a global scale, but is either the result of true differences between galaxies or the mixing of phases of the ISM on sub-kiloparsec scales.

\begin{figure*}[htp] 
   \centering
   \plotone{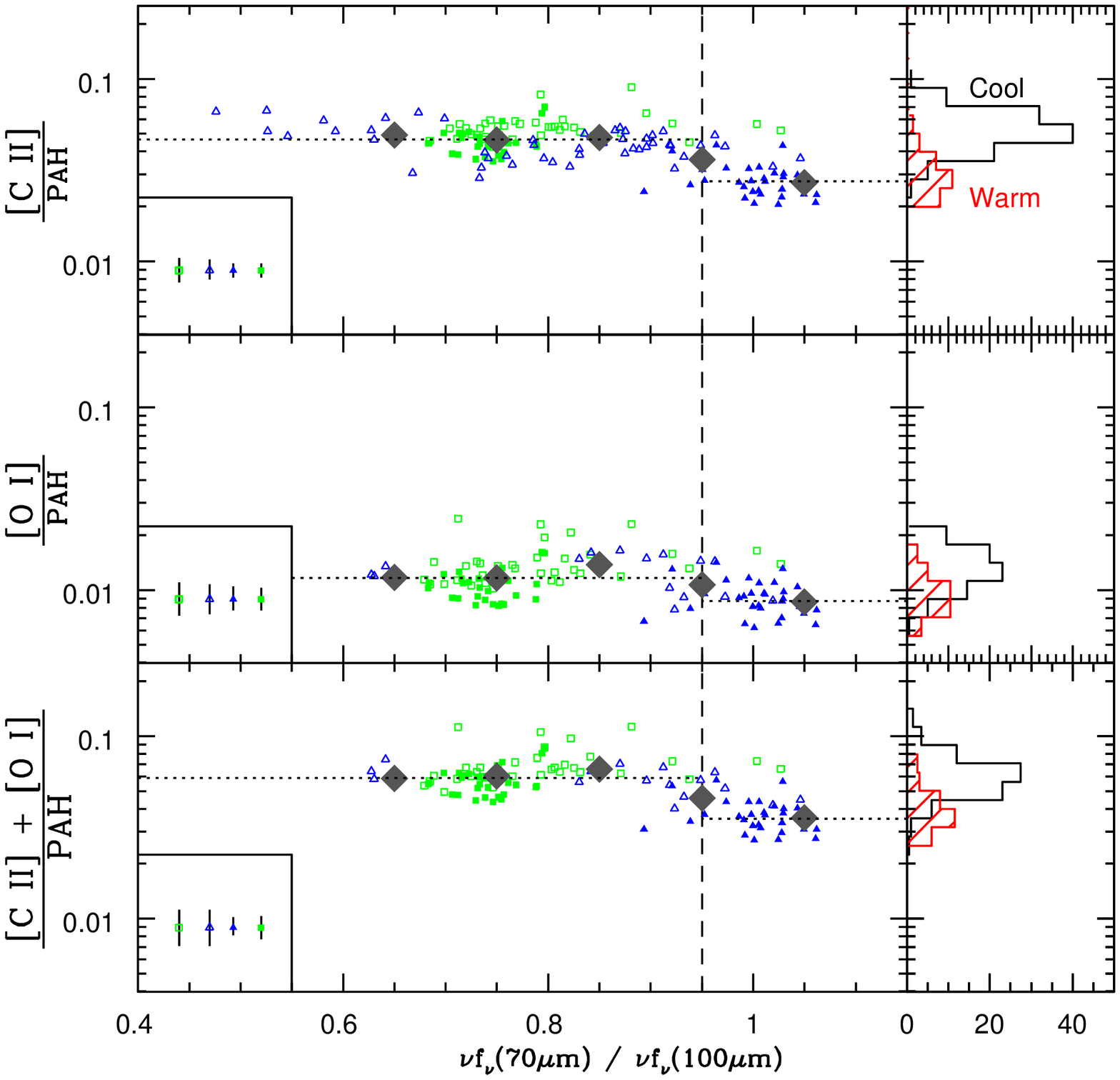}
   \caption{Top: [C~\ii] 158 $\mu$m / PAH plotted as a function of the far infrared color, $\nu f_\nu(70) / \nu f_\nu(100)$.  Middle: [O~\one] 63 $\mu$m / PAH plotted as a function of the far infrared color, $\nu f_\nu(70) / \nu f_\nu(100)$.  Bottom: ([C~\ii] 158 $\mu$m + [O~\one] 63 $\mu$m) / PAH plotted as a function of the far infrared color, $\nu f_\nu(70) / \nu f_\nu(100)$, where PAH is traced by either PAH, PAH$^*_{SL}$ or PAH$^*_{LL}$, as available.  Regions in NGC~1097 are plotted as blue triangles whereas regions from NGC~4559 are plotted as green squares.  Filled points indicate regions wherein PAH was measured in the full IRS spectrum, while open points denote estimates of PAH from Equations 3 and 4.   The large grey diamonds show the weighted mean when data are binned. Histograms show the distribution of heating intensity for regions with gas warmer (red) and cooler (black) than $\nu f_\nu(70) / \nu f_\nu(100)$=0.95.  The mean error for each data set are shown to the lower left.}
    \label{fig:lineoverpahstar}
\end{figure*}

Theoretical models of the photoelectric effect show that PAHs and other small grains dominate the production of photoelectrons in neutral gas \citep[e.g., ][]{bakes,weing}.  This prediction has been supported by the strong correlation of [C~\ii] emission and mid-IR flux in nearby galaxies.  In addition to providing photoelectrons, PAHs simultaneously emit photons in the mid-IR via vibrational and bending transitions.  These mid-IR bands should provide a more precise view of heating and cooling than integrated thermal dust emission, which includes emission from large grains that are inefficient at heating the gas in the ISM.  As PAHs are more directly coupled to heating efficiency than TIR, they may offer a more robust and precise measure of the heating efficiency, e.g.,    
\begin{equation}\frac{I_{\mbox{[C~\ii]}} + I_{\mbox{[O~\one]}}}{I_{\mbox{PAH}}} = (\eta_{\mbox{[C~\ii]}} + \eta_{\mbox{[O~\one]}})\epsilon^\prime_{\mbox{{\tiny PE}}},\end{equation} 
where $\epsilon^\prime_{\mbox{\tiny PE}}$ is the ratio of gas heating (via photoelectrons) to dust heating from PAHs, traced by emission in the mid-IR.  In the sample used by \citet{malhotra}, the globally averaged ratio of [C~\ii] and PAH emission shows remarkably little scatter \citep[$\sigma~\sim~0.16$ dex, ][]{helou}, compared to the ratio of [C~\ii] and the infrared continuum.  Indeed, they found that PAH emission between 5 and 10 $\mu$m was directly proportional to the intensity of [C~\ii] line emission. If PAHs and emission from the [C~\ii] fine-structure line are closely linked, heating by PAHs must decrease at high f$_\nu$(60 $\mu$m) / f$_\nu$(100 $\mu$m).  Thus, \citet{helou} posited that large grains become relatively more important to heating as the incident radiation field increases in intensity.

\section{Line Deficits in NGC~1097 and NGC~4559}
We show the ratio of fine-structure cooling line emission to PAH emission, as traced by either PAH, PAH$^*_{SL}$ or PAH$^*_{LL}$, as available, in Figure~\ref{fig:lineoverpahstar}.  The dispersion in ([C~\ii] + [O~\one])/PAH ($\sigma~\sim$ 0.13 dex) is reduced compared to the ([C~\ii] + [O~\one])/TIR ($\sigma~\sim$ 0.20 dex), suggesting that ([C~\ii] + [O~\one])/PAH is a more direct measure of the heating efficiency in the ISM.  When we focus specifically on the emission from small grains, neither NGC~1097 or NGC~4559 exhibit a gradual decline of ([C~\ii] + [O~\one])/PAH as a function of the far-IR color for regions described by a far-IR color cooler than $\sim$ 0.95.  We also note that the apparent offset of $\epsilon_{\mbox{{\tiny PE}}}$ between NGC~1097 and NGC~4559 has disappeared.  This lack of offset suggests that C$^+$ and PAHs are spatially correlated to a greater extent than C$^+$ and TIR or that the C$^+$ and PAH abundances are linked.  This is supported by recent Herschel results which indicate that TIR is indeed influenced by old stellar populations \citep{bendo2011}; these populations are inefficient at heating the gas, but easily heat interstellar dust.

\begin{figure}[tp] 
   \epsscale{1.2}
   \centering
   \plotone{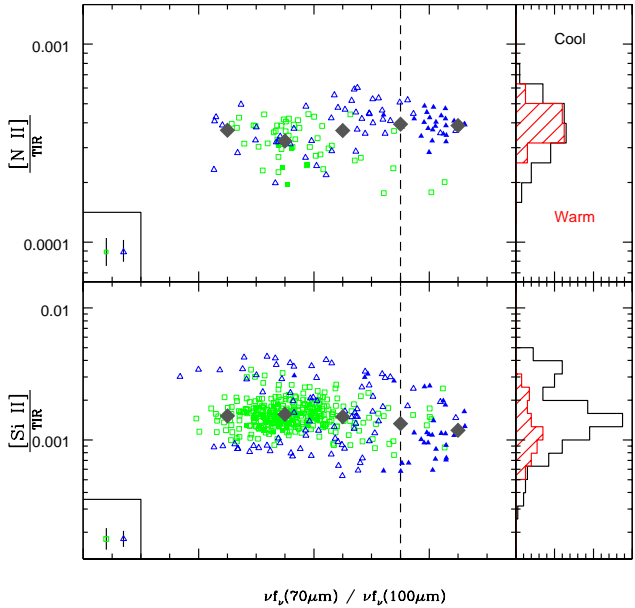}
   \caption{Top: [N~\ii] 122 $\mu$m / TIR plotted as a function of the far infrared color, $\nu f_\nu(70) / \nu f_\nu(100)$.  Bottom: [Si~\ii] 35 $\mu$m / TIR plotted as a function of the far infrared color, $\nu f_\nu(70) / \nu f_\nu(100)$.  If a dusty H~\ii\ region was the cause of the line deficit, then all lines that become optically thick should show the deficit, which is not seen in either [N~\ii] / TIR or [Si~\ii] / TIR. Colors and shapes of points follow the same scheme as Figure \ref{fig:lineoverpahstar}.}
    \label{fig:nii}
\end{figure}

Over much of the range in $\nu f_\nu$(70 $\mu$m)/$\nu f_\nu$(100 $\mu$m), the line strength of both [C~\ii] and [O~\one] are proportional to PAH emission, indicating the vibrational emission of PAH bands is directly proportional to the rate of photoelectron ejection from those same grains.  However, in regions of the ISM characterized by warm dust, $\nu f_\nu$(70 $\mu$m)/$\nu f_\nu$(100 $\mu$m) $\gtrsim$ 0.95, we see a significant drop in both [C~\ii]/PAH and [O~\one]/PAH, signaling that the correlation between the ejection of photoelectrons and vibrational excitation of PAHs is not constant over the full range of bulk dust temperature (see histograms on the right of Figure \ref{fig:lineoverpahstar}).  A value of 0.95 in the ratio $\nu f_\nu$(70 $\mu$m)/$\nu f_\nu$(100 $\mu$m) is representative of the color above which $\epsilon_{\mbox{{\tiny PE}}}$ drops and does not imply a single temperature that defines a break in the relation or physical conditions.  As such, this color indicates where a bulk change in the physical conditions is seen; we have denoted this color by the vertical line in Figures \ref{fig:mal}, \ref{fig:lineoverpahstar}, and \ref{fig:nii}.

We investigate, in these spatially resolved galaxies, scenarios that have been proposed to explain the far-IR line deficit in global studies to ascertain the cause of this drop in $\epsilon_{\mbox{{\tiny PE}}}$ at warm far-IR colors.  Notably we investigate  the likelihood that self absorption, soft radiation fields, density effects on [C~\ii] line emission, and further ionization of PAHs can explain this occurrence in NGC~1097 and NGC~4559.

\subsection{Self-absorbed optically thick [C~\ii]}
[C~\ii] and [O~\one] luminosities could be diminished if the photons were absorbed by cool foreground gas \citep{luhman}.  Self absorption has been observed in both the [O~\one] 63 $\mu$m line in Arp 220 \citep{Fischer1997} and the [C~\ii] 158 $\mu$m line in observations of warm molecular gas in the Milky Way \citep{Boreiko1997}.  However, these occurrences of self absorption were seen in highly obscured regions; Arp 220, a nearby ultraluminous infrared galaxy (ULIRG), has been shown to have unusually high optical depth \citep[$\tau_{100\micron}~\sim~5$,][]{rangwala}; while the PDR observed by \citet{Boreiko1997} was located on the far side of the molecular cloud, with a minimum possible optical depth of 0.6 in the [C~\ii] line.  

Using the ratio of H$\alpha$/H$\beta$ from \citet{kenn09} we find an average extinction, $\langle A_v\rangle$, of 1.40 and 0.50 for NGC~1097 and NGC~4559 respectively.  Looking only at the 20\arcsec$\times$20\arcsec\ circumnuclear regions of each galaxy yields slightly higher amounts of extinction, 2.34 and 2.22 \citep{Moustakas}.  However, these values are all considerably lower than the extinction seen in Arp 220 where self-absorbed [C~\ii] has been detected.  It may also be noted that the extinction in the nucleus of NGC~4559 is only 0.14 magnitudes lower than the extinction seen in the ring of NGC~1097.  Despite this similarity, regions in the nucleus of NGC~4559 are not described by the warm dust temperatures seen in regions with a line deficit.


Cold foreground diffuse clouds with $A_v$ $\sim$ 1.0 could produce an optical depth of $\sim$1.5 in the [C~\ii] line center. These clouds would have to contain cold C$^+$ at the same velocity, within a few km s$^{-1}$, as the emitting cloud.  A correlation in both velocity and position is unlikely between unassociated clouds.  A chance alignment is expected to be more likely in edge-on systems, where more clouds could lie in the line-of-sight. Contrary to this expectation, the line deficit is primarily observed in NGC 1097, a roughly face-on galaxy, and not as robustly in NGC 4559 which is inclined 64$^\circ$ to our line of sight.  This scenario becomes even less likely given that the position of these clouds must correlate with an increasing dust temperature.  

Even so, we can determine an upper limit on the amount of C$^+$ self- absorption present by using the silicate absorption features at 9.8 and 18 $\mu$m as a surrogate, since the mid-IR continuum absorbed by silicates originates in clouds that are likely to produce self-absorption.  Scaling the silicate opacity will yield only an upper limit to C$^+$ self absorption as silicate absorption is insensitive to the velocity structure of the intervening clouds. The de-convolved [C~\ii] linewidth in most regions is narrower than the PACS instrumental profile at 157 $\mu$m (FWHM = 240 km s$^{-1}$); thus, line widths in our data are dominated by the instrumental profile.  Observations of [C~\ii] emission in a galactic star-forming region and its associated PDR show widths of 5-10 km s$^{-1}$ using HIFI on {\it Herschel's} \citep{dedes}, which has a velocity resolution of 0.7 km s$^{-1}$.  For narrow [C~\ii] emission, i.e., 4 -- 10 km s$^{-1}$, the [C~\ii] line becomes optically thick at N(H) = 3.65 -- 6.25$\times10^{21}$ cm$^{-2}$ \citep{russell}, assuming $N(C^+)/N(H) = 1.6 \times 10^{-4}$.  Using the 10 $\mu$m silicate absorption profile from \citet{li}, these column densities would produce a minimum silicate opacity, $\tau_{9.7}$, of 0.135 -- 0.270.  We find marginal detections of silicate absorption in only 13 apertures where full Spitzer-IRS coverage exists. These regions, which are located on the western edge of the central ring of NGC~1097, yield $\langle\tau_{9.7}\rangle$ = 0.06, with a maximum of $\tau_{9.7} = 0.142$.  

While some self-absorption by C$^+$ is possible, the $\tau_{9.7}$ yielded by fitting Spitzer-IRS spectra is not consistent with large amounts of self-absorption exist.  We conclude that the existence of the far-IR line deficit in these galaxies is not attributable to self absorption, which should introduce, at most, a drop in ([C~\ii] + [O~\one])/PAH by a factor of two.  Nonetheless, we cannot eliminate the possibility that intervening molecular clouds may exist that increase the scatter we see in ([C~\ii] + [O~\one])/PAH and ([C~\ii] + [O~\one])/TIR.

\subsection{Soft UV radiation Field}
[C~\ii] emission in the local ISM will decrease after O and B stars have died, as photoelectons are not ejected as frequently in the softer radiation field \citep[e.g.,][]{spaans}, whereas the dust, including PAHs, continues to be heated.  The existence of a soft radiation field in the warmest parts of a galaxy seems unlikely, e.g., a high $\nu f_\nu(70) / \nu f_\nu(100)$ does not imply an absence of O and B stars.  Indeed, recent observations of Paschen $\alpha$ emission in the central ring have shown that the very regions that exhibit a line deficit have a higher star formation rate than other parts of NGC~1097 \citep{Hsieh}.  The elevated star formation and FUV flux \citep{Kuchinski} associated with the ring in NGC~1097, which is associated with many of the apertures that show a line deficit, lead us to reject a soft radiation field due to a dearth of massive stars.

\subsection{Dusty H~\ii\ Regions}
The presence of large amounts of dust in the vicinity of photoionized gas could also lead to the creation of the line deficit \citep{luhman2}.  Direct absorption of these photons by the dust also leads to an increase in TIR  while decreasing the UV photons that escape and heat the neutral gas, thus leading to a net reduction in $\epsilon_{\mbox{{\tiny PE}}}$.  The models of \citet{abel2009} have shown that the presence of dust in regions with a high ionization parameter removes Lyman continuum photons that would normally heat the gas.  The absence of these photons leads to a reduction of far-IR fine-structure lines from both the ionized and neutral gas.  However, with inclusion of radiation pressure, large grains may be compressed into an ionized shell, leading to an increase in the stellar photons that ionize H, compared to a uniform dust distribution, negating the ability of dust to cause a decrease in far-IR line emission \citep{draine2011}.

Neither [N~\ii] 122 $\mu$m /TIR nor [Si~\ii] 35 $\mu$m / TIR show a trend with far-IR color, as illustrated in Figure \ref{fig:nii}.  The lack of a deficit in the [N~\ii] line, which is only found in ionized gas, is particularly difficult to explain by invoking a dusty H~\ii\ region.  If any trend is notable in [N~\ii]/TIR it is an {\it increase} in regions characterized by warm far-IR color.  The different [N~\ii]/TIR plateaus may be related to the metallicity difference between NGC~1097 and NGC~4559, as secondary production of nitrogen may exist at the higher metallicity of NGC~1097 \citep{Costas}.  Alternatively, this effect could also be the result of H~\ii\ regions in the ring of NGC~1097 filling a larger fraction of the ISM than other regions where [N~\ii] emission is detected. 

Global studies of ULIRGS have indicated the presence of dusty regions and a deficit in the [N~\ii] 122 $\mu$m line \citep{fischer2010,Carpio}. We see no indication of this effect in resolved regions of NGC~1097 and NGC~4559.  While we cannot rule out the applicability of dusty H~\ii\ Regions to ULIRGs, we do not find evidence of elevated dust levels in the ionized gas within these two galaxies.  However, \citet{beirao2011} find $L_{FIR}/M_{H_2}$ $\sim$ 50 $L_\odot/M_\odot$ in NGC~1097.  While this is below the ratio of $L_\odot/M_\odot$ where \citet{Carpio} see a line deficit begin (80 $L_\odot/M_\odot$), it lies near this domain.  If we plot data from \citet{Carpio} with the ring of NGC~1097, we find that regions on the central ring of NGC~1097 lie at the lower edge of the scatter in their data and transition nicely into the deficit seen in galaxies with higher $L_{FIR}/M_{H_2}$.

\begin{figure}[htp] 
   \epsscale{1}
   \centering
   \plotone{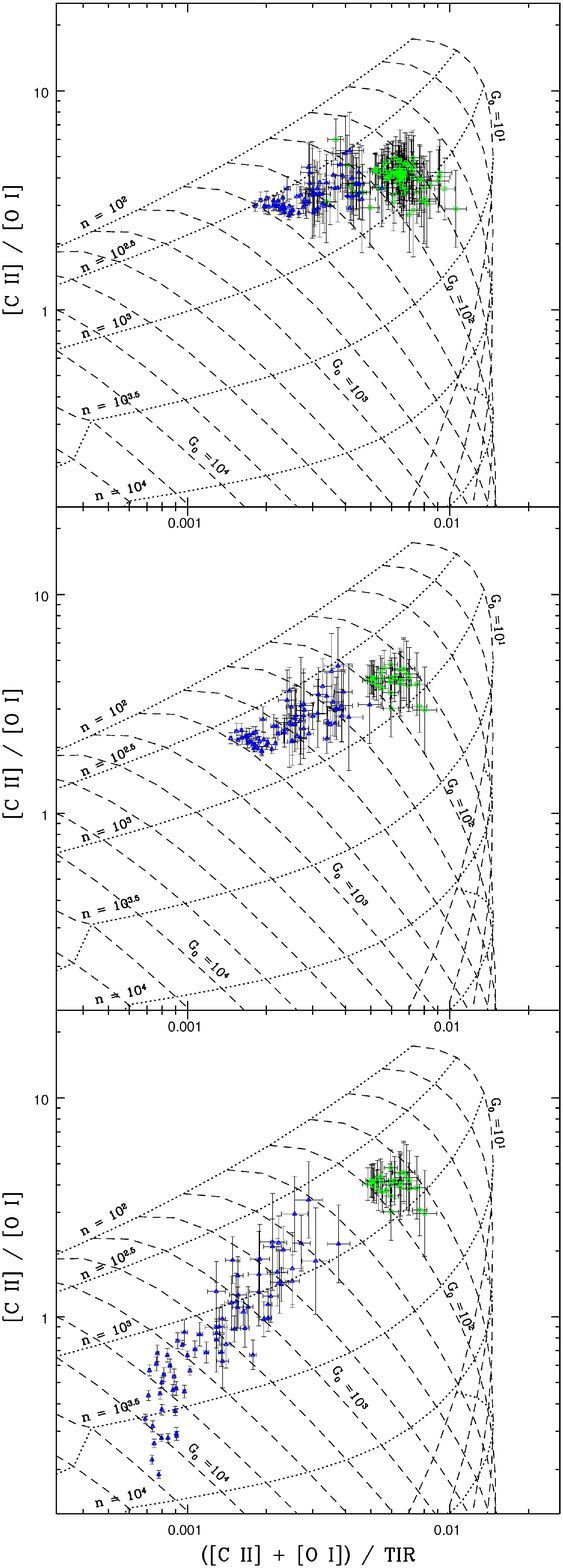}
   \caption{[C~\ii] 158 $\mu$m/[O~\one] 63 $\mu$m plotted against ([C~\ii] 158 $\mu$m + [O~\one] 63 $\mu$m)/TIR in NGC~1097 (blue triangles) and NGC~4559 (green squares).  A grid of G$_0$, in units of the local Galactic radiation field, and $n_H$, in $cm^{-3}$, as determined from the PDR models of \citet{kaufman06} are over-plotted.  Only data with a reliable [N~\ii] 122 $\mu$m line detection are plotted when correcting for an ionized component.  In the middle plot we assume a $n_e$ of 100 cm$^{-3}$ when the [S \iii] ratio is in the low-density limit.    In the bottom plot we assume a $n_e$ of 10 cm$^{-3}$ when the [S \iii] ratio is in the low-density limit.  Error bars do not include uncertainties in the correction for diffuse [C~\ii] emission.}
   \label{fig:wolf}
\end{figure}

\subsection{Contributions to [C~II] from Non-PDR Gas}
To understand the far-IR line emission, we must also understand the conditions of the ISM that lead to far-IR line emission.  We determine the average gas density, $n_H$, and radiation intensity, $G_0$ within each region of NGC~1097 and NGC~4559, assuming that all the emission originates from PDRs, by comparing [C~\ii], [O~\one], and TIR in each aperture to models of ionization from PDRs \citep{kaufman06}.  Data are overlaid on plots of model [C~\ii] / [O~\one]  versus ([C~\ii] + [O~\one]) / TIR in Figure~\ref{fig:wolf}.  

According to PDR models, the gas which dominates the cooling in NGC~1097 and NGC~4559 lies at densities between 10$^{2.5}$ and 10$^3$ cm$^{-3}$.  If the line deficit is caused by density effects, the [O~\one] line should become the dominant cooling line in the densest clouds.  To the contrary, as shown in Figure \ref{fig:lineoverpahstar}, rather than becoming the dominant line, [O~\one] exhibits a deficit congruent with the [C~\ii] deficit.  This indicates that the observed line deficit is not a product of increased gas density.  

The total integrated [C~\ii] emission along a line of sight includes contributions from ionized gas in addition to the PDRs.  This additional [C~\ii] emission must be removed, as the models of \citet{kaufman06} only account for emission from neutral gas.  The [C~\ii] contribution from ionized gas can be traced by emission from N$^+$, which in the Galaxy is emitted primarily in the diffuse ionized phase \citep{bennet94}.  

\begin{figure}[t] 
   \epsscale{1.2}
   \centering
   \plotone{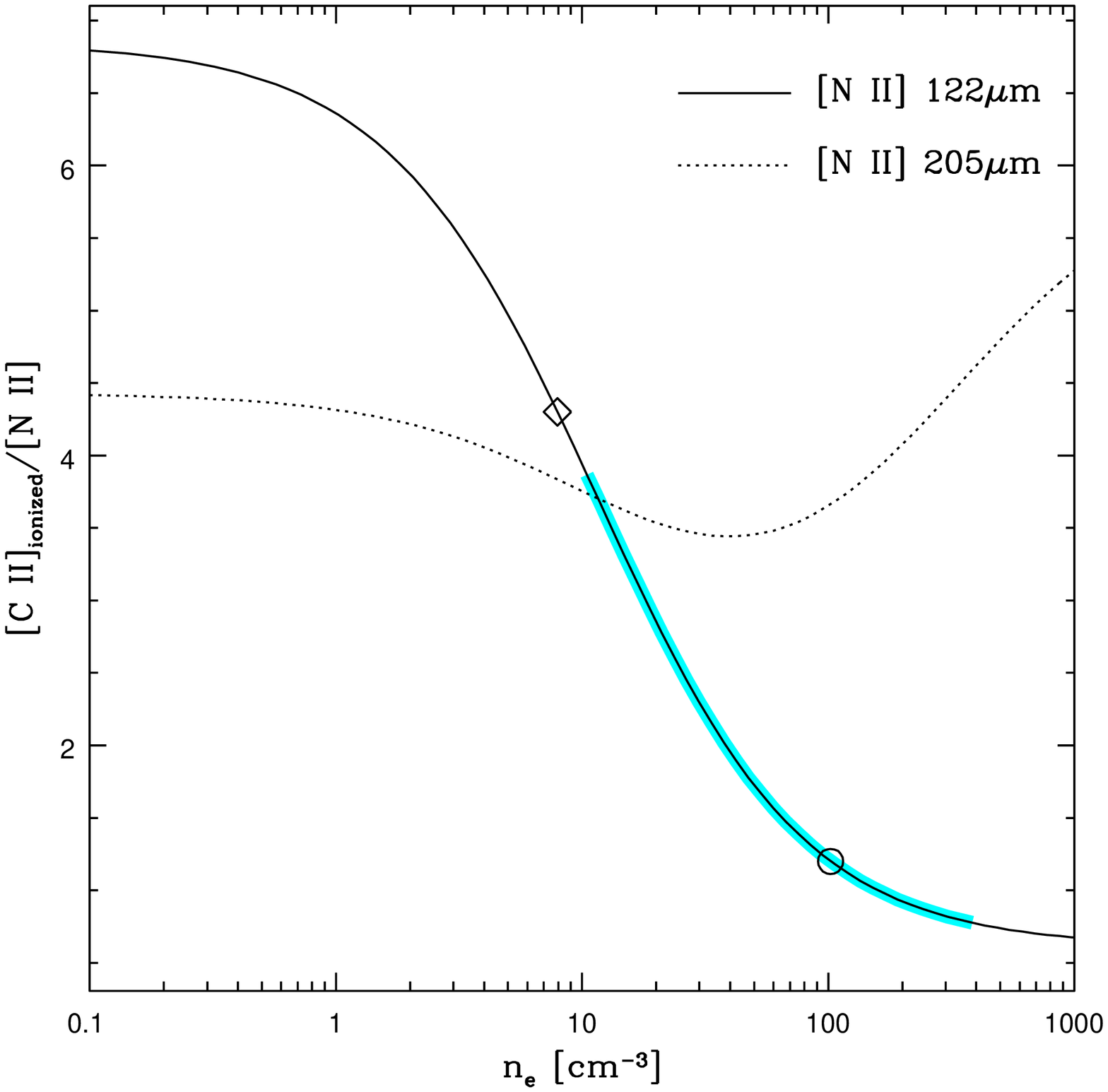}
   \caption{The ratio of [C~\ii] / [N~\ii] from diffuse ionized gas is plotted as a function of $n_e$ for both the 122 $\mu$m and 205 $\mu$m transitions.  The range of [C~\ii] / [N~\ii] 122 $\mu$m we expect to see in NGC~1097 and NGC~4559, based on density variations, is highlighted by the cyan strip.  For comparison, the single value correction used by \citet{malhotra} is shown as a diamond on the curve tracing [C~\ii] / [N~\ii] 122 $\mu$m; whereas the circle denotes the [C~\ii] / [N~\ii] 122 $\mu$m determined from the [N~\ii] 122 $\mu$m/[N~\ii] 205 $\mu$m ratio by \citet{beirao2011}.}
   \label{fig:ciinii}
\end{figure}

This correction is straightforward given measurements of [N~\ii] 205 $\mu$m, as this faint line has the same critical density as the [C~\ii] 158 $\mu$m line in ionized gas ($\sim$ 45 cm$^{-3}$) \citep{oberst}.  This results in a [C~\ii] 158 $\mu$m/[N~\ii] 205 $\mu$m ratio that is nearly constant (see Figure \ref{fig:ciinii}), varying by only a factor of 1.6 over three orders of magnitude in density \citep{oberst}.  Unfortunately, the faint [N~\ii] line at 205 $\mu$m is difficult to detect.  The only observations of this line for these two galaxies cover the star forming ring in NGC~1097.  

\citet{beirao2011} employ the region in which [N~\ii] 205 $\mu$m emission is observed to determine that emission from diffuse ionized carbon is equal to $\sim$1.2 $\times$ [N\ii] 122 $\mu$m.  As we are interested in studying emission extended well beyond the brightest regions we must allow for this ratio to change. There are no observations of [N~\ii] 205 $\mu$m outside the nucleus of NGC~1097 that allow us to track these changes.  Furthermore, no observations of [N~\ii] 205 $\mu$m exist to verify that this ratio is the same in NGC~4559, which has a lower star formation rate \citep{calzetti}

An alternative approach is to use the relatively stronger [N~\ii] 122 $\mu$m line as a means of tracing the diffuse ionized [C~\ii] emission \citep[e.g.,][and references therein]{kaufman99,malhotra}.  The higher critical density of [N~\ii] 122 $\mu$m ($\sim$ 300 cm$^{-3}$) leads to a stronger dependence of  [C~\ii] 158 $\mu$m/[N~\ii] 122 $\mu$m on density compared to [CII] 158 um/[NII] 205 um.  This is shown in Figure \ref{fig:ciinii} which plots the [C~\ii]/[N~\ii] line ratios as a function of electron density.  Preferably, this density would be calculated from the ratio of [N~\ii] 122 $\mu$m/[N~\ii] 205 $\mu$m but as discussed, the weaker line is unobserved for most of this sample.  A substitute density indicator is available in the mid-IR ratio of [S \iii] 18.71 $\mu$m /[S \iii] 33.48 $\mu$m.  

\begin{figure}[htp] 
   \centering
   \epsscale{1.2}
   \plotone{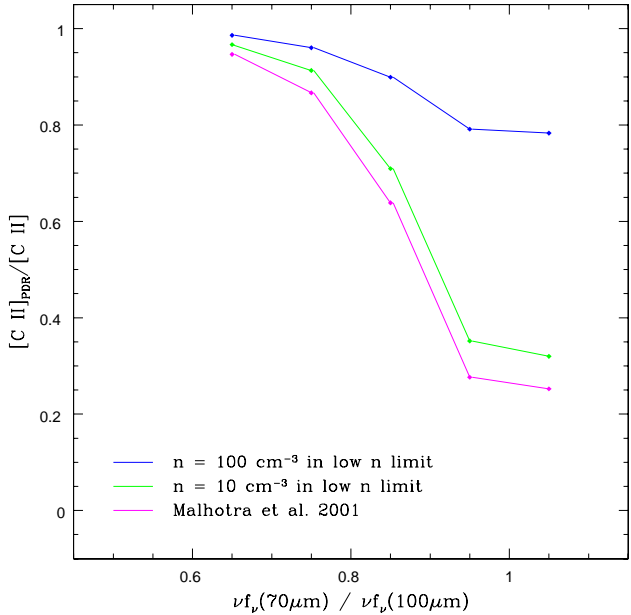}
   \caption{The fraction of [C~\ii] emission attributed to PDRs is shown as a function of $\nu f_\nu(70) / \nu f_\nu(100)$.  All methods of correcting for emission from diffuse ionized gas show a decrease in the PDR fraction at warm colors.  Rather than eliminating the line deficit, a plot of [C~\ii]$_{PDR}$ / PAH would significantly increase the line deficit as the fraction of [C~\ii] emanating from PDRs decreases in warm regions.}
   \label{fig:frac}
\end{figure}

The [S \iii] ratio is sightly temperature dependent and is most sensitive to densities in the range 100 -- 10$^4$ cm$^{-3}$ \citep{snidjers}.  We measure this ratio in our LL  coverage of NGC~1097 and NGC~4559, when these lines are detected, and derive an electron density, $n_e$, using the models of \citet{smith09}.  We derive [C~\ii] 158 $\mu$m/[N~\ii] 122 $\mu$m appropriate for ionized gas as a function of $n_e$, in order to scale the [N~\ii] emission appropriately.  To determine this ratio, we used observed Galactic gas phase abundances of C \citep[1.6$\times$10$^{-4}$ per hydrogen nucleus,][]{sofia} and N \citep[7.5$\times$10$^{-5}$ per hydrogen nucleus,][]{meyer}. For nitrogen we adopted the transition probabilities from \citet{galavis} and  collision strengths from \citet{hudson}; while for carbon we use transition probabilities from \citet{galavis2} and collision strengths from \citet{blum}.  

The correction for diffused ionized gas using the [S~\iii] lines to trace the density is in agreement with the correction that would be adopted using the measurements of [C~\ii]/[N~\ii] 205 $\mu$m in the nucleus of NGC~1097, see Figure \ref{fig:ciinii}.  We show the magnitude of this correction visually by plotting our data on models of PDR emission both with and without correcting for diffuse ionized gas in Figure \ref{fig:wolf}.  There is a significant shift in 20\% - 40\% of the data points with a 3$\sigma$ detection of [N~\ii] 122 $\mu$m, depending on assumptions in the low density limit, when emission from ionized gas is removed.  Given the uncertainties in correcting for contributions to [C~\ii] from ionized gas, reality most likely lies between these representations of the UV field and the gas density we show in Figure \ref{fig:wolf}.  

When the [S \iii] ratio is in the low-density limit, $n_e$ $\lesssim$ 100 cm$^{-3}$, the ratio of [C~\ii]/[N~\ii] 122$\mu$m in ionized gas can vary between 1.24, at $n_e$ of 100 cm$^{-3}$, and 6.8, at $n_e$ of 0.1 cm$^{-3}$.  Observations of the [N~\ii] 205 $\mu$m line \citep{beirao2011} indicate that at least some of the regions that are in the low density limit have an $n_e$ of $\sim$ 100 cm$^{-3}$; therefore, in the nucleus of NGC~1097 we adopt the conservative scale factor of 1.24 appropriate for $n_e$ = 100 cm$^{-3}$.  In regions outside the central star forming ring of NGC~1097, the flux of the [N~\ii] $122\mu$m line is sufficiently weak relative to [C~\ii] that the effects of this uncertainty are negligible (e.g., \textless 5\%).

If observed PAH emission only receives contributions from the diffuse ionized gas, then $\epsilon^\prime_{\mbox{{\tiny PE}}}$ should be calculated using only [C~\ii] emission that originates from PDRs.  We show the fraction of [C~\ii] emission that emanates from PDRs as a function of the far-IR color in Figure \ref{fig:frac}.  In addition to the correction adopted in this work, we also calculate the fraction of [C~\ii] that would originate in PDRs using the single value prescription from \citet{malhotra}.  It is noteworthy that all regions show a relatively high PDR fraction, regardless of the far-IR color.  Furthermore, the fraction of [C~\ii] emitted by the PDR decreases in the warmer regions, regardless of which correction for ionized gas is employed.  An $\epsilon^\prime_{\mbox{{\tiny PE}}}$ associated with pure PDR emission would thus exhibit an \emph{increased} line deficit at warm far-IR colors, accentuating the sudden decrease in ([C~\ii] + [O~\one]) / PAH.

\begin{figure}[htp] 
   \epsscale{1.2}
   \centering
   \plotone{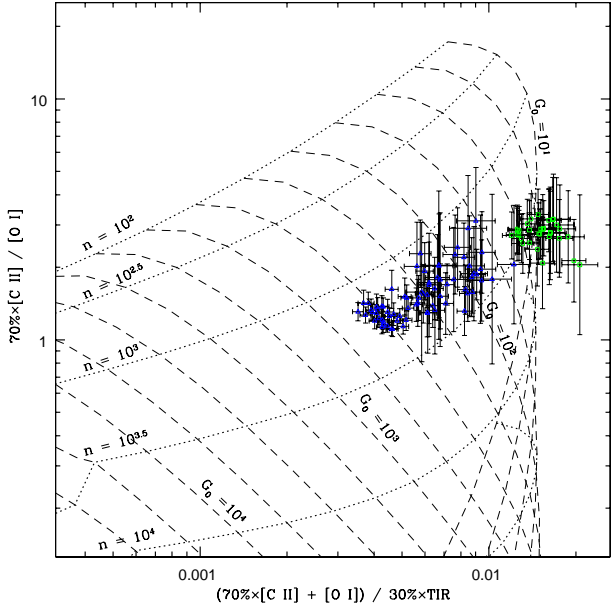}
   \caption{Similar to the middle panel of Figure \ref{fig:wolf} except that we assume 70\% of TIR and 30\% of [C~\ii] emission originates in diffuse clouds.}
   \label{fig:wolfdiff}
\end{figure}

In addition to an ionized component and PDRs, the ISM also contains diffuse HI.  Observations of the dust SED in these galaxies \citep{aniano2} are incompatible with the high $G_0$ we determine assuming that all the [C~\ii] emission and TIR originates in a PDR with an implied starlight intensity heating, $\langle U \rangle$, that is inconsistent, typically 10 -- 50$\times$ lower than expected from $G_0$.  In reality, some fraction of both TIR and the [C~\ii] line will emerge from the diffuse ISM.  Attributing 70\% of TIR and 30\% of [C~\ii] surface brightness to cool diffuse gas would bring the $G_0$ and from PDR model and $\langle U \rangle$ from the dust SED into closer agreement, typically $\langle U \rangle$ 1 -- 10$\times$ lower than expected from $G_0$.  This will simultaneously increase the $n_H$ of the modeled PDR, as shown in Figure \ref{fig:wolfdiff}.  Standard models of starlight heating indicate that this diffuse gas is too cool to emit [O~\one].  Decreasing the amount of [C~\ii] relative to [O~\one] emitted by the PDR could result in [C~\ii] no longer being the dominant coolant.  However, the inclusion of other line coolants, e.g., [O~\one], does not eliminate the line deficit.  It is noteworthy that models of starlight heated diffuse gas also fail to produce the warm $H_2$ emission seen in some translucent clouds \citep{ingalls}.  H$_2$ emission lines are seen in our {\it Spitzer} IRS spectra, indicating that the diffuse cloud population may also harbor hot material that supplements the [O~\one] emission, requiring non-standard heating processes that are not fully described by a PDR model.  

Assuming that diffuse gas is distributed in a relatively uniform manner throughout the ISM, correcting TIR, [C~\ii], and [O~\one] for its contributions would alter the absolute values of $n_e$ and $G_0$, \emph{ but would not change the general trend of points relative to each other}, see Figure \ref{fig:wolfdiff}.  PDR models are not able to reproduce all the spectral line and dust SED features we have observed.  However, none of the corrections we have investigated would alter the fact that the ratio of $G_0/n_e$ and $\langle U \rangle$ both increase in regions where the line deficit is observed.  Contributions of line emission and TIR from diffuse gas, both ionized and cool, cannot produce the far-IR line deficit in NGC~1097 and NGC~4559.  

\begin{figure}[bp] 
   \centering
   \epsscale{1.2}
   \plotone{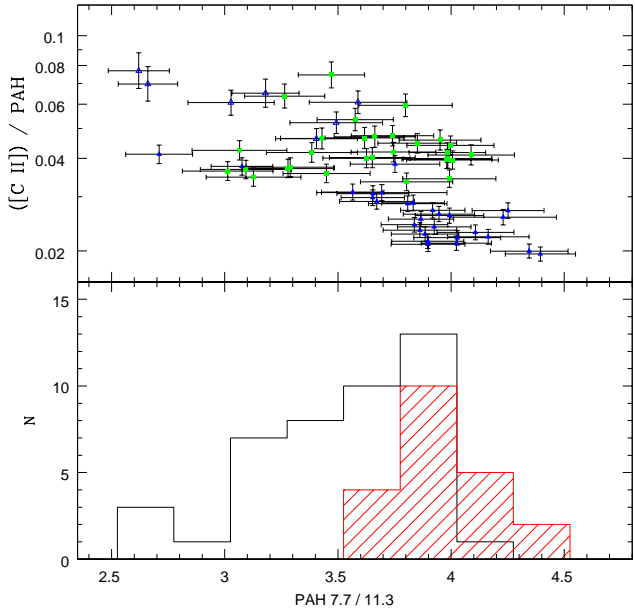}
   \caption{Top: [C~\ii]/PAH is plotted as a function of the PAH 7.7 / 11.3 $\mu$m ratio.  An increase in this PAH ratio, which is observed in all regions showing a deficit, is indicative of an increase in PAH ionization.  Data shown here are limited to where SL observation had been acquired, see Figure \ref{fig:aors}.  Colors and shapes of points follow the same scheme as Figure \ref{fig:lineoverpahstar}. Bottom: Histogram of the PAH 7.7 / 11.3 $\mu$m ratio. Regions for which $\nu f_\nu$(70 $\mu$m)/$\nu f_\nu$(100 $\mu$m) \textless\ 0.95 are indicated with the black histogram, whereas regions with $\nu f_\nu$(70 $\mu$m)/$\nu f_\nu$(100 $\mu$m) \textgreater\ 0.95 are indicated with the red histogram.}
   \label{fig:histo}
\end{figure}

\newpage\subsection{Grain Charging}
Positively charged dust grains are inefficient at heating gas due to the greater potential well that must be overcome to eject additional electrons \citep{TH85}.  This decrease in gas heating is not accompanied by a strong decrease in the thermal dust emission that dominates TIR; nor does it cause the vibrational PAH emission to drop in strength \citep{draine}.  Therefore, an increase in grain ionization will lead to a deficit in line cooling relative to the thermal dust and PAH emission \citep{tielens1999}.  

Figure \ref{fig:wolf} shows that most of the gas in NGC~1097 and NGC~4559 lies at a density between 10$^{2.5}$ and 10$^3$~cm$^{-3}$.  However, the local UV field, $G_0$, varies considerably, ranging from 50 - 1000, when the most conservative corrections for ionized gas are employed.  We see the largest variation of UV heating in the ring of NGC~1097, while NGC~4559 displays remarkable uniformity in both $n_H$ and G$_0$ across its disk.  We compare ([C~\ii] + [O~\one])/PAH  and ([C~\ii] + [O~\one])/TIR to $G_0/n_H$ derived from models of PDRs and the mean dust-weighted starlight intensity, $\langle U\rangle$, derived from dust models of the IR SED \citep{draine}, where the average value of $\langle U\rangle$ in each region was extracted from maps produced by \citet{aniano2}.  Both gas and dust diagnostics clearly indicate that strong radiation fields are incident on the regions showing a line deficit.  Indeed, $G_0/n_H$ increases more than an order of magnitude between regions of low and high incident radiation, suggesting that grain charging could be important in these warm regions \citep{wolfire1990}. 

If the ionization of PAHs is indeed responsible for the observed line deficit, a significant variation in the 7.7 $\mu$m to 11.3 $\mu$m PAH band flux ratio should also be observed \citep{draine}.  The 7.7 and 11.3 $\mu$m PAH features are both observed using the SL module and have very similar PSFs, which are considerably smaller than the 12\arcsec\ apertures we have used.  As the convolution of Spitzer-IRS maps are very uncertain \citep{Pereira-Santaella} we compare the 7.7 $\mu$m to 11.3 $\mu$m PAH band flux ratio measured from un-convolved spectra with the same ratio measured from convolved spectra.  As expected, convolution of SL spectra decreases the spread seen in this PAH band ratio.  The un-convolved spectra clearly show that more ionized PAHs are found in regions where the line-to-PAH deficit is strongest.  

In Figure \ref{fig:histo} we plot convolved [C~\ii]/PAH as a function of the ratio of the un-convolved 7.7 $\mu$m/11.3 $\mu$m PAH bands.  Regions that show a far-IR fine-structure line deficit relative to total PAH emission do indeed exhibit a modestly greater than average 7.7 $\mu$m / 11.3 $\mu$m ratio, suggesting that PAH ionization fraction is higher in regions with a line deficit.  This is consistent with an elevated $G_0/n_H$ seen in these regions. 

The shift seen in the 7.7 $\mu$m to 11.3 $\mu$m PAH band flux ratio could also be indicative of a change in the grain size distribution.  Small PAHs are more efficient at producing photoelectrons, compared to the larger PAH grains, as they tend to stay neutral rather than becoming charged \citep{bakes}.  This has been confirmed via Galactic observations of the Horsehead nebula that show a smaller PAH/VSG (very small grain) ratio in the warm gas compared to the diffuse ISM \citep{Compiegne}.  However, we do not detect a correlation between either the 7.7 $\mu$m/17 $\mu$m nor the 11.3 $\mu$m/17 $\mu$m PAH band ratio and ([C~\ii] + [O~\one])/PAH.  If grain size was the driving factor in the suppression of  ([C~\ii] + [O~\one])/PAH, we should detect a significant trend with both of these ratios as the 17 $\mu$m complex is attributed to the largest PAHs.  Furthermore, we find that ([C~\ii] + [O~\one])/PAH does not correlate with the 6.2/7.7 $\mu$m band ratio, which has traditionally been associated with grain size \citep{draine01}.  


\section {Conclusions}
In this paper we have sought to understand the physical conditions that drive the heating and cooling of gas and dust in the ISM of two nearby galaxies using spectral mapping data of NGC~1097 and NGC~4559, which are among the first galaxies observed by the PACS spectrometer on the {\it Herschel Space Observatory} as part of the {\sc Kingfish} program.  We matched these maps to {\sc Sings} mid-IR observations to extract spatially resolved information on both gas and dust heating in these two nearby galaxies.  Using these combined observations we have investigated photoelectric heating efficiency and the importance of PAHs as a source of photoelectrons over a large range in surface brightness with higher spatial resolution than previously possible in external galaxies.

As seen in other galaxies, we find that the photoelectric heating efficiency, as traced by ([C~\ii] + [O~\one])/TIR, is lower where the dust temperature is elevated.  Contrary to a decreasing heating efficiency, which is evident in NGC~1097, ([C~\ii] + [O~\one])/PAH is flat as the temperature of the dust increases, up to $\nu f_\nu$(70 $\mu$m)/$\nu f_\nu$(100 $\mu$m) $\sim$ 0.95.  In this regime, gas heating, as traced by fine-structure cooling line emission, and dust heating, discerned by the power emitted in the mid-IR from PAHs, effectively trace each other, indicating that the ratio of ([C~\ii] + [O~\one])/PAH is uniform in many environments.  The scatter observed in ([C~\ii] + [O~\one])/PAH is also reduced compared to ([C~\ii] + [O~\one])/TIR.  This suggests that PAH emission delineates the ejection of photoelectrons more effectively than the integrated dust emission, TIR.  In gas with dust warmer than $\nu f_\nu$(70 $\mu$m)/$\nu f_\nu$(100 $\mu$m) $\sim$ 0.95, however, we observe a deficit in [C~\ii]/PAH and [O~\one]/PAH, that was not previously detected.  

Many explanations have been proposed to explain the deficit in the observed heating efficiency. Comparing the properties of regions in which this deficit is seen in NGC~1097 and NGC~4559 we find that:
\begin{itemize}
\item Both [C~\ii] and [O~\one] exhibit line deficits relative to both TIR and PAH emission.  However, deficits are not observed in the [N~\ii] or [Si~\ii] lines, as would be expected if dusty H\ii\ regions were responsible.
\item If the line deficit was the result of increased density, cooling by [O~\one] should compensate for the dearth of [C~\ii] emission in dense gas.  This is not the case, as [O~\one] and [C~\ii] exhibit similar deficits.
\item Correcting [C~\ii] emission by removing contributions to this line from both ionized and diffuse gas, does not remove the line deficit.  To the contrary, this correction would accentuate the [C~\ii]/PAH deficit observed in regions with warm far-IR colors.
\item Both $\langle U\rangle$ from SED models and the ratio of $G_0/n_H$ from models of PDRs indicate that regions where a deficit is seen have intense radiation fields that could lead to grain charging.
\item The distribution of the 7.7 $\mu$m/11.3 $\mu$m PAH band strength ratio differs between warm regions, that exhibit a line deficit, and cool regions that do not.  This indicates that grain charging, which leads to inefficient photoelectric heat, could be responsible for the observed line deficits. 
\end{itemize}

While NGC~1097 and NGC~4559 have regions spanning a large range of IR surface brightness, they represent a small fraction of the diverse environments probed by {\sc Kingfish}.  An analysis of the heating and cooling in the full sample will probe more intense starbursts, larger ranges of metallicity, and different types of ionization sources.  By analyzing the full {\sc Kingfish} sample we will be able to more fully study the range of gas and dust heating explored here and expand our investigation to even more extreme environments to test the universality of these results. 

\acknowledgments
PACS has been developed by a consortium of institutes led by MPE (Germany) and including UVIE (Austria); KU Leuven, CSL, IMEC (Belgium); CEA, LAM (France); MPIA (Germany); INAF-IFSI/OAA/OAP/OAT, LENS, SISSA (Italy); IAC (Spain). This development has been supported by the funding agencies BMVIT (Austria), ESA-PRODEX (Belgium), CEA/CNES (France), DLR (Germany), ASI/INAF (Italy), and CICYT/MCYT (Spain).  
HIPE is a joint development by the Herschel Science Ground Segment Consortium, consisting of ESA, the NASA Herschel Science Center, and the HIFI, PACS and SPIRE consortia.  
This work is based [in part] on observations made with Herschel, a European Space Agency Cornerstone Mission with significant participation by NASA. Support for this work was provided by NASA through an award issued by JPL/Caltech.  
This research has made use of the NASA/IPAC Extragalactic Database (NED) which is operated by the Jet Propulsion Laboratory, California Institute of Technology, under contract with the National Aeronautics and Space Administration.


\clearpage

\begin{deluxetable}{lcccccccccc}  
\tabletypesize{\scriptsize}
\tablecaption{PACS Spectroscopic Observations of NGC 1097}
\tablewidth{0pt}
\tablehead{   
  \colhead{Obs. ID}	&
  \colhead{R.A.}	&
  \colhead{Dec.}	&
  \colhead{Mode\tablenotemark{1}}	&
  \colhead{Raster}	&
  \colhead{Raster}	&
  \colhead{Duration}	&
  \colhead{Cycles}	&
  \colhead{Repe-}	&
  \colhead{On-Source}	&
  \colhead{Line}	\\
  \colhead{}		&
  \colhead{(J2000)}	&
  \colhead{(J2000)}	&
  \colhead{}	&
  \colhead{Pattern}&
  \colhead{Step (\arcsec)}	&
  \colhead{(s)}	&
  \colhead{}	&
  \colhead{titions}	&
  \colhead{(s raster$^{-1}$)}	&
  \colhead{}}
\startdata
1342189609	& 02 46 14 &	-30 17 00	&	WS		&	2x2	&	18x47&	675		&	1		&	2	&	320		&	[CII] 157.7		\\
1342189608	& 02 46 12 &	-30 17 10	&	CN		&	2x1	&	12x12&	13200	&	2		&	8	&	6144	&	[OI] 63.18		\\
1342189415	& 02 46 14 &	-30 15 04	&	WS		&	2x2	&	12x12&	1644	&	1		&	1	&	160		&	[CII] 157.7		\\
		& $\prime\prime$ & $\prime\prime$ & $\prime\prime$ &	$\prime\prime$ & $\prime\prime$ & $\prime\prime$ &$\prime\prime$&	3	&	480		&	[NII] 121.9		\\
1342189412	& 02 46 23 &	-30 17 50	&	WS		&	2x2	&	12x12&	758		&	1		&	1	&	160		&	[CII] 157.7		\\
1342189411	& 02 46 19 &	-30 16 28	&	WS		&	2x2	&	12x12&	2262	&	1		&	1	&	160		&	[CII] 157.7		\\
		& $\prime\prime$ & $\prime\prime$ & $\prime\prime$ & $\prime\prime$ & $\prime\prime$ & $\prime\prime$ &$\prime\prime$&	1	&	160		&	[NII] 121.9		\\
1342189410	& 02 46 19 &	-30 16 28	&	CN		&	2x2	&	12x12&	13833	&	1		&	1	&	704		&	[CII] 157.7		\\
		& $\prime\prime$ & $\prime\prime$ & $\prime\prime$ & $\prime\prime$ & $\prime\prime$ & $\prime\prime$ &$\prime\prime$&	1	&	704		&	[OI] 63.18		\\
		& $\prime\prime$ & $\prime\prime$ & $\prime\prime$ & $\prime\prime$ & $\prime\prime$ & $\prime\prime$ &$\prime\prime$&	2	&	1408	&	[NII] 121.9		\\
1342189409	& 02 46 23 &	-30 17 50	&	WS		&	2x2	&	12x12&	1217	&	1		&	4	&	640		&	[NII] 121.9		\\
1342189278\tablenotemark{2}&02 46 23 &-30 15 57.2&WS& 	2x2 &	18x47&	2570	& 	1		& 	9	&	1440	&	[NII] 121.9		\\
1342189277	& 02 46 23 &	-30 15 57.2	&	WS		&	2x2	&	12x47&	945		&	1		&	3	&	480		&	[CII] 157.7		\\
1342189074	& 02 46 19 &	-30 16 29.6	&	CN		&	2x1	&	12x12&	7459	&	1		&	2	&	688		&	[CII] 157.7		\\
		& $\prime\prime$ & $\prime\prime$ & $\prime\prime$ & $\prime\prime$ & $\prime\prime$ & $\prime\prime$ &$\prime\prime$&	8	&	2752	&	[NII] 121.9		\\
1342189073	& 02 46 12 &	-30 17 10.7	&	CN		&	2x1	&	12x12&	7459	&	1		&	2	&	688		&	[CII] 157.7		\\
		& $\prime\prime$ & $\prime\prime$ & $\prime\prime$ & $\prime\prime$ & $\prime\prime$ & $\prime\prime$ &$\prime\prime$&	8	&	2752	&	[NII] 121.9		\\
1342188939	& 02 46 14 &	-30 17 00.8	&	WS		&	2x2	&	18x47&	2568	&	1		&	9	&	1440	&	[NII] 121.9		\\
1342188938	& 02 46 23 &	-30 17 50.5	&	CN		&	2x2	&	12x12&	7903	&	1		&	1	&	704		&	[CII] 157.7		\\
		& $\prime\prime$ & $\prime\prime$ & $\prime\prime$ & $\prime\prime$ & $\prime\prime$ & $\prime\prime$ &$\prime\prime$&	1	&	704		&	[OI] 63.18		\\
		& $\prime\prime$ & $\prime\prime$ & $\prime\prime$ & $\prime\prime$ & $\prime\prime$ & $\prime\prime$ &$\prime\prime$&	3	&	2112	&	[NII] 121.9		\\
1342188936	& 02 46 15 &	-30 16 48.8	&	CN		&	2x1	&	12x12&	13200	&	2		&	8	&	6144	&	[OI] 63.18	\enddata
\label{t:obs1097}
\tablenotetext{1}{CN = Chop-Nod;  WS = Wavelength Switching}
\tablenotetext{2}{Parts of these observations were affected by grating malfunctions.  Affected raster positions have not been included in combined maps. }
\tablecomments{Summary of PACS spectral observations of NGC~1097.  Units of right ascension are hours, minutes, and seconds, and units of declination are degrees, arcminutes, and arcseconds. When multiple lines were grouped into an observation each line is listed on a separate line; only observational parameters that are line specific are given in additional lines of an observation.}
\end{deluxetable}

\begin{deluxetable}{lcccccccccc}  
\tabletypesize{\scriptsize}
\tablecaption{PACS Spectroscopic Observations of NGC 4559}
\tablewidth{0pt}
\tablehead{   
  \colhead{Obs. ID}	&
  \colhead{R.A.}	&
  \colhead{Dec.}	&
  \colhead{Mode\tablenotemark{1}}	&
  \colhead{Raster}	&
  \colhead{Raster}	&
  \colhead{Duration}	&
  \colhead{Cycles}	&
  \colhead{Repe-}	&
  \colhead{On-Source}	&
  \colhead{Line}	\\
  \colhead{}		&
  \colhead{(J2000)}	&
  \colhead{(J2000)}	&
  \colhead{}	&
  \colhead{Pattern}&
  \colhead{Step (\arcsec)}	&
  \colhead{(s)}	&
  \colhead{}	&
  \colhead{titions}	&
  \colhead{(s raster$^{-1}$)}	&
  \colhead{}}
\startdata
1342187229 & 12 36 02 & +27 56 26.9 					& CN &	2x1& 16x14.5&6674 	&	1&	8&	1536&	[O I] 63.18		\\
1342187230 & 12 36 02 & +27 56 26.9 					& CN &	2x1& 24x22  &3851 	&	1&	4&	688	&	[N II] 121.9	\\
		& $\prime\prime$ & $\prime\prime$ & $\prime\prime$ & $\prime\prime$ & $\prime\prime$ & $\prime\prime$ &$\prime\prime$&	1&	172	&	[C II] 157.7	\\
1342187231 & 12 36 00 & +27 57 02.9 					& CN &	2x1& 16x14.5&6674 	&	1&	8&	1536&	[O I] 63.18		\\
1342187232 & 12 36 00 & +27 57 02.9 					& CN &	2x1& 24x22  &3851 	&	1&	4&	688	&	[N II] 121.9	\\
		& $\prime\prime$ & $\prime\prime$ & $\prime\prime$ & $\prime\prime$ & $\prime\prime$ & $\prime\prime$ &$\prime\prime$&	1&	172	&	[C II] 157.7	\\
1342187233 & 12 35 57 & +27 57 36.0 					& CN &	2x2& 16x14.5&6866 	&	1&	4&	768	&	[O I] 63.18		\\
1342187235 & 12 35 57 & +27 57 36.0 					& CN &	2x2& 24x22  &4829 	&	1&	2&	344	&	[N II] 121.9	\\
		& $\prime\prime$ & $\prime\prime$ & $\prime\prime$ & $\prime\prime$ & $\prime\prime$ & $\prime\prime$ &$\prime\prime$&	1&	172	&	[C II] 157.7	\\
1342189062 & 12 35 57 & +27 57 36.0 					& WS &	2x2& 12x12  &396	&	1&	1& 	40	&	[C II] 157.7	\\
1342189063 & 12 35 57 & +27 57 36.0 					& WS &	2x2& 12x12  &2559 	&	1&	9&	360	&	[N II] 121.9	\\
1342189066\tablenotemark{2}	& 12 36 01 & +27 56 44.1	& WS &	2x2& 18x47  &2917	&	1&	8&	320	&	[N II] 121.9 	\\
		& $\prime\prime$ & $\prime\prime$ & $\prime\prime$ & $\prime\prime$ & $\prime\prime$ & $\prime\prime$ &$\prime\prime$&	2&	80	& 	[C II] 157.7	\\
1342189068 & 12 35 52 & +27 58 45.2 					& WS &	2x3& 18x47  &997	&	2&	1&	80	&	[C II] 157.7	\\
1342189069\tablenotemark{2}	& 12 35 52 & +27 58 45.2	& WS &	2x3& 18x47  &3410	&	1& 	9&	360 & 	[N II] 121.9
	
\enddata
\label{t:obs4559}
\tablenotetext{1}{CN = Chop-Nod;  WS = Wavelength Switching}
\tablenotetext{2}{Parts of these observations were affected by grating malfunctions.  Affected raster positions have not been included in combined maps. }
\tablecomments{Summary of PACS spectral observations of NGC~4559.  Units of right ascension are hours, minutes, and seconds, and units of declination are degrees, arcminutes, and arcseconds. When multiple lines were grouped into an observation each line is listed on a separate line; only observational parameters that are line specific are given in additional lines of an observation.}
\end{deluxetable}




\end{document}